\documentclass[letterpaper,twocolumn,10pt]{article}
\usepackage{usenix-2020-09}

\usepackage{amsmath}
\usepackage{graphicx}
\usepackage{booktabs}
\usepackage{tabularx}
\usepackage{caption}
\usepackage{placeins}
\usepackage{stfloats}
\usepackage{xurl}
\usepackage{tikz}
\usetikzlibrary{shapes.geometric}
\usepackage{pgfplots}
\pgfplotsset{compat=1.18}
\usepackage{listings}

\providecommand{\Description}[2][]{}

\lstdefinestyle{chainseed}{language=Python,basicstyle=\ttfamily\footnotesize,
  columns=fullflexible,keepspaces=true,showstringspaces=false,
  keywordstyle=\color{violet!75!black},
  stringstyle=\color{teal!75!black},commentstyle=\color{black!55},
  emph={EthereumService,createBlockchain,createBeaconSetupNode,
    createGethNode,createBeaconNode,createVcNode,connectToGethNode,
    connectToBeaconNode,setBootNode,enablePOSValidatorAtGenesis,
    setGenesis,addLocalAccount,createFaucetServer,MoneroService,SolanaService,
    createSeedNode,createClientNode,enableWallet,
    createBootstrapValidator,createValidator},
  emphstyle=\color{blue!65!black},
  breaklines=true,frame=single,rulecolor=\color{black!30},
  backgroundcolor=\color{black!2},captionpos=b}

\newcommand{\floatnote}[1]{%
  \par\smallskip{\normalfont\footnotesize\raggedright\noindent
  \textit{Note.} #1\par}%
}

\setcitestyle{numbers,square,sort&compress}

\hypersetup{
  hidelinks,
  pdftitle={ChainSEED: A Scalable, High-Fidelity Blockchain Emulator},
  pdfauthor={Qishu Jin, Wenliang Du, Kairui Feng},
  pdfkeywords={Blockchain emulation, performance evaluation, scalability, fidelity}
}

\title{ChainSEED: A Scalable, High-Fidelity Blockchain Emulator}
\author{
{\upshape Qishu Jin}\\
Zhejiang University\\
Hangzhou, China\\
{\small\texttt{\href{mailto:brucejin@zju.edu.cn}{brucejin@zju.edu.cn}}}
\and
{\upshape Wenliang Du}\\
Zhejiang University\\
Hangzhou, China\\
{\small\texttt{\href{mailto:wenliangdu@zju.edu.cn}{wenliangdu@zju.edu.cn}}}
\and
{\upshape Kairui Feng}\\
Zhejiang University\\
Hangzhou, China\\
{\small\texttt{\href{mailto:fkrcarry@zju.edu.cn}{fkrcarry@zju.edu.cn}}}
}
\date{}

\begin{document}
\pagestyle{plain}
\raggedbottom
\maketitle

\begin{abstract}
A blockchain is a shared ledger maintained by networked nodes through a consensus protocol. It enables participants to maintain a common record without relying on a single ledger operator and provides a foundation for digital assets and programmable applications. To understand its performance and security and evaluate new mechanisms and applications, researchers need experimental environments with configurable workloads, controllable network conditions, and observable outcomes. However, public blockchains and testnets limit experimental control, model-based simulators omit real client implementations, and real-client environments face challenges in combining network control with resource-efficient deployment.

In this paper, we present ChainSEED, a scalable, high-fidelity blockchain emulator built on the SEED Internet Emulator. It separates blockchain configuration from client placement on a configurable routed network. For Ethereum PoS, the service organizes separately configured execution, consensus, and validator clients and provides APIs for customizing genesis state, peer connections, and client software. A transaction generator informed by Ethereum mainnet activity, application interfaces, and chain- and node-level monitoring support experimental interaction and observation. In addition to Ethereum, we also developed blockchain emulators for the Monero and Solana blockchains.

To support larger experiments within limited hardware budgets, we examine beacon-node memory reduction, execution-client memory allocation, and validator-client multiplexing, together with their effects on transaction processing and consensus. Using our optimization method, we are able to deploy 400 beacon nodes, 400 execution nodes, and 60,000 validator identities on a single server with approximately 1~TiB of memory and 384 logical CPU cores, sustaining approximately 20~TPS. We are also able to emulate a 1,200-node Monero network, configure mining experiments with up to 200 active single-threaded PoW miners, and emulate a 200-node Solana network with sustained finalization. We demonstrate ChainSEED's research utility via several case studies, including routing attacks, blockchain amplification attacks, Monero eclipse attacks, topology-discovery benchmarking against known peer graphs, and network-configuration experiments.
\end{abstract}
\noindent\textbf{Keywords:} Blockchain emulation, performance evaluation, scalability, fidelity.

\section{Introduction}
\label{sec:introduction}
A blockchain is a shared ledger maintained by networked nodes. Transactions are grouped into blocks linked to their predecessors, forming a history of recorded activity. Nodes exchange and validate these records, while a consensus protocol determines which blocks enter the ledger. This allows participants to maintain a common record without a single ledger operator. Beyond digital-asset transfers, platforms such as Ethereum support smart contracts, programs that apply predefined rules to update blockchain state, enabling applications such as digital payments, asset exchanges, and lending~\cite{ethereum-blockchain-intro}.

Research on blockchain systems spans consensus and scaling mechanisms~\cite{cblocksim2022}, security and privacy~\cite{doumanidis2026routing,deanonymizing-validators-2025}, decentralization and economic incentives~\cite{deanonymizing-validators-2025,ethereum-incentive-research}, and smart-contract and cross-chain applications~\cite{blockbench2017,paraswap2025}. Researchers propose new mechanisms and investigate the behaviors and security weaknesses of existing deployments. Evaluating a mechanism requires examining its intended benefits and trade-offs. Investigating an existing system may involve reproducing a phenomenon, identifying its cause, or testing a suspected vulnerability and defenses. Both benefit from an experimental environment in which researchers can configure operating conditions, observe behaviors, test their conclusions, and repeat experiments under comparable conditions.

Public blockchains support studies of deployed systems and real activity, while testnets allow application testing without spending mainnet assets~\cite{ethereum-networks}. These shared environments give individual researchers limited control over network participants and background activity. Experiments comparing network conditions, reproducing failures, or evaluating attacks and defenses require control over participating nodes and their interactions. Isolated deployments let researchers vary these conditions without disrupting other users.

A private test network provides a starting point by allowing researchers to deploy a blockchain independently of public networks~\cite{ethereum-networks}. More broadly, blockchain simulation and emulation platforms provide configurable environments for experimental research~\cite{blocksim2020,ethshadow,containerchain2021,kurtosis-ethereum}. Making these environments useful presents three challenges:
\begin{enumerate}
\item \textbf{Network conditions and workloads.} Experiments need configurable network conditions and transaction workloads that reflect the scenario under study. The challenge is to provide this control while preserving the characteristics relevant to the investigation.
\item \textbf{Interaction and observation.} Researchers need to integrate their software, interact with the experimental system, and collect and visualize its behavior. These capabilities must work together to connect experimental actions with observable outcomes.
\item \textbf{Scalability.} Limited memory and CPU budgets constrain experimental scale. The challenge is to support larger networks within these budgets while maintaining timely transaction processing and continued consensus progress.
\end{enumerate}

To address these challenges, we present \emph{ChainSEED}, a scalable, high-fidelity blockchain emulator built on the SEED Internet Emulator~\cite{du2022seed}. ChainSEED separates blockchain configuration from client placement in the emulated network. Its Ethereum service supports customization of genesis state, peer connections, and client software, complemented by a configurable transaction generator, application interfaces, and chain- and node-level monitoring. For Ethereum PoS, independent configuration of execution, consensus, and validator clients allows us to investigate resource and operational trade-offs when scaling node networks, transaction loads, and validator populations. We extend the same service architecture to Monero and Solana. This paper makes three contributions:
\begin{itemize}
\item We design and implement ChainSEED, organizing real blockchain clients as configurable services on SEED's emulated network. It provides APIs for chain and client customization, tools for workload generation, application interaction, and monitoring, and support for Ethereum, Monero, and Solana.
\item Using our optimization method, we are able to deploy 400 beacon nodes, 400 execution nodes, and 60,000 validator identities on a single server with approximately 1~TiB of memory and 384 logical CPU cores, sustaining approximately 20~TPS. We are also able to emulate a 1,200-node Monero network with continued block production and transaction inclusion and configure mining experiments with up to 200 active single-threaded PoW miners. We further emulate a 200-node Solana network with sustained finalization.
\item We demonstrate ChainSEED's research utility via several case studies, including security studies, topology-discovery benchmarking against known peer graphs, and network-configuration experiments.
\end{itemize}

Upon acceptance, all ChainSEED system code will be released on GitHub.


\section{Background and Related Work}
\label{sec:background}
\subsection{Ethereum PoS Blockchain}
\label{sec:ethereum-architecture}
\begin{figure*}[!tb]
\centering
\resizebox{\linewidth}{!}{
\begin{tikzpicture}[x=1cm,y=1cm,font=\sffamily\footnotesize,
  role/.style={draw=black!55,rounded corners=3pt,
    minimum width=3.7cm,minimum height=1.6cm},
  title/.style={align=center,text width=3.5cm,
    font=\sffamily\footnotesize\bfseries,inner sep=0pt},
  task/.style={align=center,text width=3.4cm,inner sep=0pt},
  peers/.style={draw=black!40,rounded corners=3pt,
    minimum width=3.5cm,minimum height=0.5cm,align=center},
  relation/.style={<->,>=stealth,line width=0.8pt,black!65}]
\node[role,fill=orange!5] (execution) at (0,0) {};
\node[role,fill=blue!4] (consensus) at (5.3,0) {};
\node[role,fill=green!5] (validator) at (10.6,0) {};

\node[title] at (0,0.48) {Execution client};
\node[task] at (0,-0.16) {Execute transactions\\Maintain account state};
\node[title] at (5.3,0.48) {Consensus (beacon) client};
\node[task] at (5.3,-0.16) {Validate blocks\\Track consensus and finality};
\node[title] at (10.6,0.48) {Validator client};
\node[task] at (10.6,-0.16) {Sign block proposals\\Submit attestations};

\draw[relation] (execution.east) -- (consensus.west);
\draw[relation] (consensus.east) -- (validator.west);

\node[peers,fill=orange!2] (executionpeers) at (0,-1.85)
  {Execution peers\\\textbf{Bootnodes} and other nodes};
\node[peers,fill=blue!2] (consensuspeers) at (5.3,-1.85)
  {Consensus peers\\\textbf{Bootnodes} and other nodes};
\draw[relation] (execution.south) -- node[right=3pt,font=\sffamily\scriptsize]
  {P2P} (executionpeers.north);
\draw[relation] (consensus.south) -- node[right=3pt,font=\sffamily\scriptsize]
  {P2P} (consensuspeers.north);
\end{tikzpicture}}
\caption{Ethereum client roles and their relationships.}
\Description{Three horizontal boxes show execution clients executing transactions and maintaining account state, consensus or beacon clients validating blocks and tracking consensus and finality, and validator clients signing block proposals and submitting attestations. Bidirectional arrows connect execution to consensus nodes and consensus to validator nodes. Separate P2P arrows connect execution and consensus nodes to peers of the same role, including bootnodes and other nodes.}
\label{fig:ethereum-node-roles}
\end{figure*}
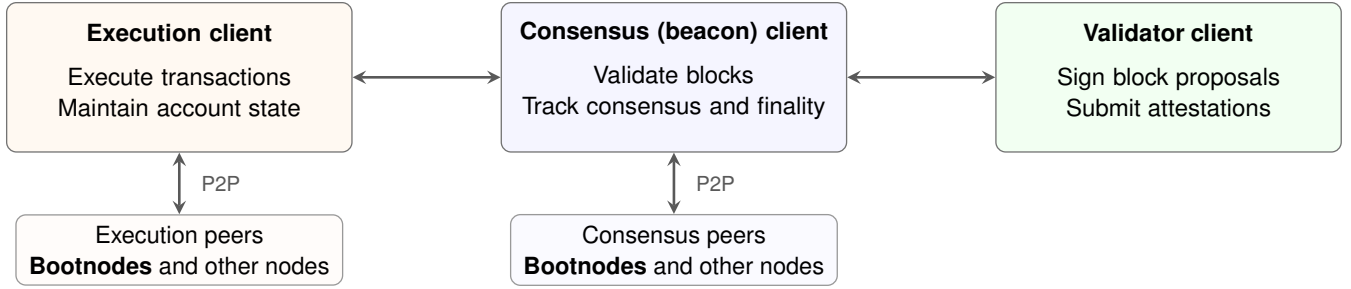

Ethereum maintains its PoS blockchain through the cooperation of three client roles: execution, consensus, and validation, as illustrated in Figure~\ref{fig:ethereum-node-roles}. Execution clients process transactions, execute smart contracts, and maintain account and contract state. Consensus clients, also called beacon clients, exchange blocks and attestations with their peers, validate blocks, and track chain selection and finality. Validator clients use signing keys to propose blocks and submit attestations through their associated beacon nodes~\cite{ethereum-node-architecture,ethereum-pos-background}. Such a design separates transaction execution from consensus while allowing the three roles to cooperate in maintaining the chain.

To operate on the same blockchain, these clients need a shared starting state, compatible protocol settings, and connections to other participants. The genesis state defines this starting point, including initial execution-side account allocations~\cite{geth-private-network}. Execution and consensus clients establish their respective P2P connections using bootstrap peers as initial contacts for node discovery~\cite{ethereum-node-architecture,ethereum-bootnodes}.

Once connected, the clients work together to process transactions and maintain the chain. A transaction submitted through a wallet enters an execution client's transaction pool and propagates to its peers. When a validator is selected to propose a block, its beacon node obtains an execution payload containing transactions and related execution block data from the associated execution client, and the validator signs the resulting block proposal. Other validators validate the proposed block, while validators submit attestations that contribute to chain selection. Finality is established through stake-weighted votes~\cite{ethereum-node-architecture,ethereum-pos-background}. Through these exchanges, nodes keep their local views of the chain synchronized.

\subsection{The SEED Internet Emulator}
\label{sec:seed-network}
The SEED Internet Emulator is an open-source platform for constructing Internet environments that run real routing software and applications in containers~\cite{du2022seed}. Its design separates network construction, routing configuration, and service deployment into reusable layers. The base layer defines autonomous systems (ASes), hosts, routers, and their connections. Routing layers configure communication within and between ASes, while service layers deploy applications such as DNS, PKI, and CDN. Users compose these layers through a Python interface, allowing network environments and application services to be configured independently and reused in different experiments~\cite{zeng2021seedemu}.

Within this architecture, a service specifies its logical nodes' software and configuration, including user-supplied binaries and files. Each node has a symbolic \emph{virtual-node name} rather than a fixed network location. Bindings map these names to base-layer hosts, allowing the same service definition to be deployed on different hosts~\cite{du2022seed,zeng2021seedemu}.

SEED then turns the composed layers into a runnable environment through rendering and compilation. Rendering resolves virtual-node bindings and integrates the layers into network and node configurations~\cite{du2022seed}. Compilation then generates deployment files for backends such as Docker or Kubernetes, through which the network and its services are launched.

\subsection{Comparison with Related Work}
Experiments help researchers test blockchain designs and understand how changing conditions affect performance and security. A good blockchain emulator should therefore have two essential features:

First, a good blockchain emulator should run real blockchain client software. Specifications describe intended behavior, but implementations determine how nodes process workloads, synchronize state, and respond to attacks. Studying these properties requires exercising deployed blockchain software. BlockSim, SimBlock, and CBlockSim model blockchain operation and propagation~\cite{faria2019blocksim,blocksim2020,simblock2019,cblocksim2022}. Ganache executes transactions and smart contracts in a local Ethereum simulator, rather than a distributed network of production execution and consensus clients~\cite{ganache}. These platforms support protocol exploration or application testing, but cannot directly expose production-client behavior. ChainSEED builds its experiments around real blockchain client software.

Second, a good blockchain emulator should provide controllable infrastructure running real routing protocols. Public blockchain nodes communicate across independently managed networks. Routing policies, link failures, and recovery alter paths and reachability, affecting block propagation, synchronization, and consensus. Researchers need to examine these mechanisms under normal and disrupted conditions. Delays or disconnections alone do not capture how routing decisions create and resolve them. Lilith uses Kollaps to model end-to-end conditions, while Ethshadow runs clients over Shadow's simulated network~\cite{lilith2025,ethshadow,shadow-network-graph}. Kurtosis's Ethereum package uses a bridge network with LAN-like connectivity in its default Docker deployment~\cite{kurtosis-ethereum,kurtosis-docker-network}. Their documented deployments do not incorporate inter-AS routing protocols. ChainSEED integrates blockchain services with SEED's Internet infrastructure for joint studies of routing and blockchain behavior.

Other frameworks also run real clients in emulated networks, but differ in deployment cost and extensibility. Proteus's VM-based deployment incurs higher per-node operating-system overhead than ChainSEED's container-based deployment, and its demonstrated Ethereum integration is limited to PoW~\cite{proteus2019}. Containerchain couples client startup with topology configuration, requiring deployment-script changes to support different blockchain architectures~\cite{containerchain2021}. These costs matter when researchers need to change blockchain architectures or enlarge experiments within limited budgets. Pennino and Pizzonia primarily address OS-level scalability challenges and demonstrate their solutions through Ethereum emulation~\cite{pennino2024scalable}. Addressing infrastructure limits alone, however, does not explain how blockchain-client configurations affect resource consumption and chain operation. ChainSEED considers OS-level resource constraints and further investigates blockchain-client optimizations, combining them with reusable container-based services for multiple blockchains.

\section{System Design}
\label{sec:design}
ChainSEED follows SEED's service design to organize real blockchain clients into configurable
chains and deploy them on emulated network hosts. We use Ethereum's PoS architecture introduced
in Section~\ref{sec:ethereum-architecture} as a reference to explain how chain instances and
client roles are organized, customized, and supported.

\subsection{ChainSEED Architecture}
\label{sec:chainseed-architecture}

\subsubsection{High-level System Architecture}
\label{sec:software-architecture}
\begin{figure*}[!tb]
\centering
\resizebox{\linewidth}{!}{
\begin{tikzpicture}[x=1cm,y=1cm,font=\sffamily\normalsize,
 panel/.style={draw=black!55,rounded corners=3pt,line width=0.8pt},
 logical/.style={circle,draw=blue!65!black,fill=white,
   minimum size=0.28cm,inner sep=0pt,line width=0.8pt},
 host/.style={draw=black!55,fill=white,minimum width=0.27cm,
   minimum height=0.34cm,inner sep=0pt,line width=0.8pt},
 flow/.style={->,>=stealth,line width=1.1pt,draw=black!70},
 binding/.style={->,>=stealth,dashed,line width=0.85pt,draw=blue!65!black},
 item/.style={panel,fill=white,align=center,inner sep=5pt}]
\path[use as bounding box] (-0.2,0.4) rectangle (18.1,7.15);
\draw[panel,draw=blue!60!black,fill=blue!2,line width=1.1pt]
 (-0.18,4.35) rectangle (8.93,7.12);
\node[font=\sffamily\large\bfseries,text=blue!50!black] at (4.375,6.82)
 {ChainSEED Service};
\foreach \left/\name/\shade in {
 0.05/Ethereum 1/blue!8,2.25/Ethereum 2/blue!8,
 4.45/Monero 1/orange!10,6.65/Solana 1/green!8}{
 \draw[panel,fill=\shade] (\left,4.6) rectangle (\left+2.05,6.35);
 \node[font=\sffamily\normalsize\bfseries] at (\left+1.025,5.97) {\name};
}
\draw[panel,fill=blue!3] (0.05,0.6) rectangle (8.7,2.65);
\node[font=\sffamily\large\bfseries] at (4.375,1.45) {Shared SEED network};
\node[text=black!70] at (4.375,0.96) {Hosts, routers, and configurable topology};
\foreach \group in {0,1,2,3}{
 \foreach \member in {0,1,2}{
  \pgfmathsetmacro{\nodeX}{0.52+2.2*\group+0.55*\member}
  \node[logical] (v-\group-\member) at (\nodeX,5.18) {};
  \node[host] (h-\group-\member) at (\nodeX,2.12) {};
  \draw[binding] (v-\group-\member.south) -- (h-\group-\member.north);
 }
}
\node[fill=white,inner sep=4pt,text=blue!65!black] at (4.4,3.52)
 {Virtual-node bindings};
\node[item,minimum width=3.15cm,minimum height=1.1cm] (compile) at (11.2,1.35)
 {\textbf{Compilation}\\[4pt]Docker / Kubernetes};
\draw[flow] (8.7,1.35) -- (compile.west);
\node[item,minimum width=3.25cm,minimum height=1.8cm,fill=black!3]
 (execution) at (15.5,1.35) {};
\node[font=\sffamily\normalsize\bfseries] at (15.5,1.92) {Execution};
\foreach \name/\x in {e1/14.55,e2/15.02,e3/15.98,e4/16.45}{
 \node[host,minimum width=0.18cm,minimum height=0.23cm]
  (\name) at (\x,1.44) {};
}
\node[circle,draw=black!60,fill=white,minimum size=0.24cm,inner sep=0pt]
 (router-a) at (14.92,0.87) {};
\node[circle,draw=black!60,fill=white,minimum size=0.24cm,inner sep=0pt]
 (router-b) at (16.08,0.87) {};
\foreach \a/\b in {e1/router-a,e2/router-a,e3/router-b,e4/router-b,router-a/router-b}{
 \draw[black!55,line width=0.6pt] (\a) -- (\b);
}
\draw[flow] (compile.east) -- (execution.west);
\draw[panel,draw=teal!65!black,fill=teal!3,line width=1.1pt]
 (9.2,2.75) rectangle (18.0,7.12);
\node[font=\sffamily\large\bfseries,text=teal!50!black] at (13.6,6.82) {Interaction};
\draw[panel,fill=white] (9.45,3.0) rectangle (14.7,6.47);
\node[font=\sffamily\normalsize\bfseries] at (12.075,6.18) {Monitoring System};
\draw[panel,draw=blue!45!black,fill=blue!4]
 (9.7,4.0) rectangle (14.45,5.86);
\node[font=\sffamily\normalsize\bfseries] at (12.075,5.56) {Visualization};
\node at (12.075,4.85) {EthExplorer};
\node at (12.075,4.3) {Ethereum Lookingglass};
\node[item,draw=blue!45!black,fill=blue!4,minimum width=4.75cm,
 minimum height=0.6cm] at (12.075,3.45) {Scripts};
\draw[panel,fill=white] (15.0,3.0) rectangle (17.75,6.47);
\node[font=\sffamily\normalsize\bfseries] at (16.375,6.18) {Applications};
\foreach \y/\name in {5.35/MetaMask,4.6/Remix,3.85/Hardhat}{
 \node at (16.375,\y) {\name};
}
\draw[<->,>=stealth,line width=1.1pt,draw=teal!60!black]
 (execution.north) -- (15.5,2.75);
\end{tikzpicture}}
\caption{Blockchain services bound to a shared SEED network for compilation, execution, and interaction.}
\Description{A ChainSEED Service box contains four boxes named Ethereum 1, Ethereum 2, Monero 1, and Solana 1, each with three logical nodes. Individual dashed bindings map these nodes to hosts in one shared SEED network below. An arrow leads to Compilation with Docker and Kubernetes, then to a compact Execution box containing hosts and routers. Above execution, an Interaction panel contains Applications with MetaMask, Remix, and Hardhat, and a Monitoring System containing Visualization with EthExplorer and Ethereum Lookingglass, and Scripts.}
\label{fig:chainseed-software-architecture}
\end{figure*}
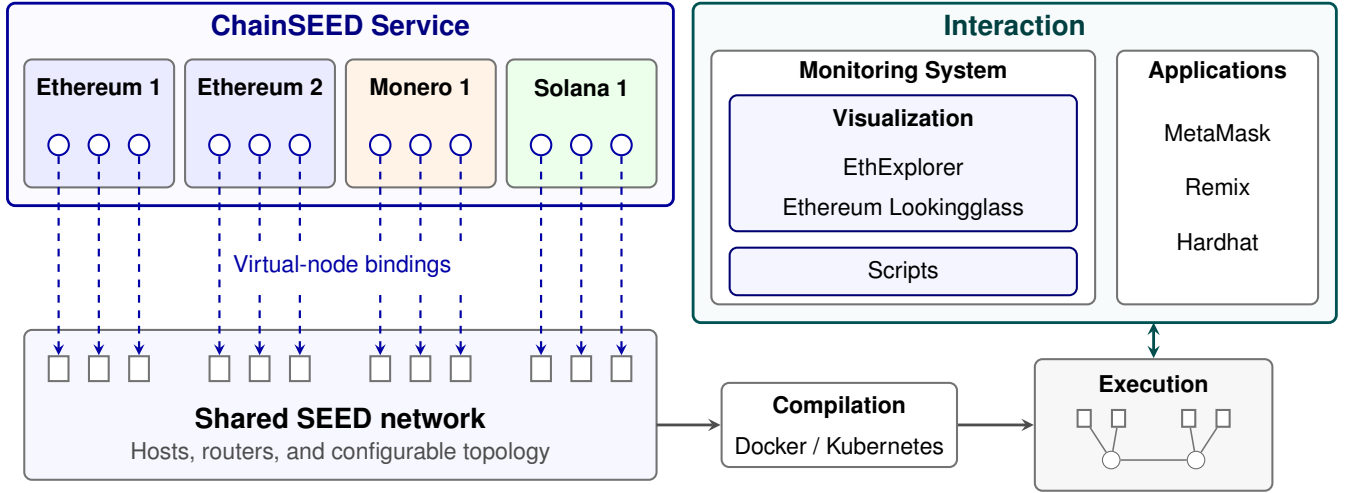

\begin{figure*}[!tb]
\centering
\input{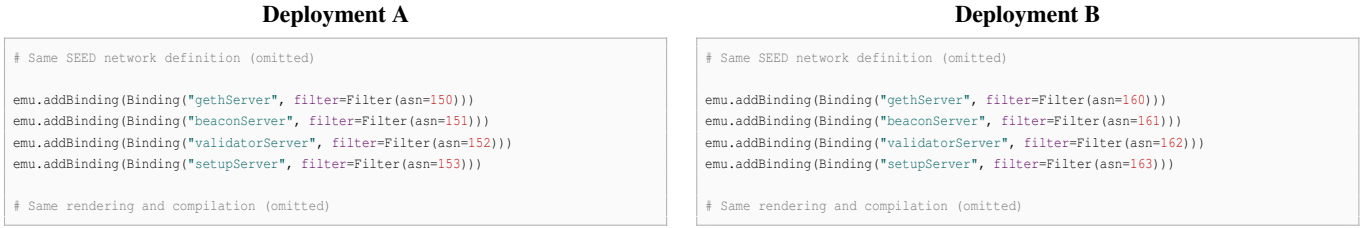}
\caption{Alternative bindings of the same chain components within one SEED network.}
\Description{Two code panels titled Deployment A and Deployment B bind the same gethServer, beaconServer, validatorServer, and setupServer names to hosts in ASes 150 through 153 or ASes 160 through 163, respectively. Both deployments use the same SEED network. Comments before and after the binding statements indicate the same omitted network definition and rendering and compilation steps. Red numbers highlight the AS choices, while all component names remain the same.}
\label{fig:service-network-separation}
\end{figure*}

ChainSEED builds on top of the SEED Internet emulator: ChainSEED constructs the
blockchain as a service, while SEED constructs the underlying network. As
illustrated in Figure~\ref{fig:chainseed-software-architecture}, the blockchain service
provides and configures the blockchain nodes, while SEED provides and configures the hosts,
and routing. They are constructed independently. Nodes in the blockchain service are logical
nodes (called virtual nodes), which will be bound to the physical nodes (hosts) defined in SEED later at the deployment phase. This late binding mechanism decouples blockchain and the underlying network, making the blockchain service portable. Figure~\ref{fig:service-network-separation} shows the binding mechanisms, as well as showing how a blockchain can be easily moved from one part of the network to another part: changing bindings is all we need. 

After binding, during rendering, the service uses bound hosts' information to complete the configuration of the blockchain system. Finally, the compilation phase generates Docker or Kubernetes deployment files. At runtime, users can interact with the blockchain emulator using the existing applications; they can also observe the chain and node behavior using our visualization and monitoring tools. 

\subsubsection{Ethereum Service Architecture}
\label{sec:blockchain-service}
\begin{figure*}[!tb]
\centering
\resizebox{\linewidth}{!}{
\begin{tikzpicture}[x=1cm,y=1cm,font=\sffamily\normalsize,
 unit/.style={draw=black!60,rounded corners=3pt,align=center,
   fill=blue!4,inner sep=7pt,line width=0.8pt},
 branch/.style={draw=black!60,line width=0.9pt},
 setting/.style={ellipse,draw=blue!55!black,fill=white,align=center,
   minimum width=3.7cm,minimum height=0.62cm,inner sep=3pt,
   text height=1.8ex,text depth=0.4ex,
   line width=0.7pt},
 role/.style={unit,minimum width=3.3cm,minimum height=1.05cm}]
\path[use as bounding box] (-0.05,-5.6) rectangle (17.1,0.5);
\node[unit,fill=blue!12,minimum width=4.8cm,minimum height=0.9cm]
 (service) at (8.5,0) {\textbf{Ethereum service}};
\node[unit,minimum width=9.1cm,minimum height=2.5cm]
 (chain) at (6.3,-2.2) {};
\node[font=\sffamily\normalsize\bfseries] at (6.3,-1.26)
 {Blockchain instance};
\node[setting] at (4.05,-1.95) {Protocol settings};
\node[setting] at (8.55,-1.95) {Genesis state};
\node[setting] at (4.05,-2.83) {Accounts};
\node[setting] at (8.55,-2.83) {Initial validators};
\node[unit,text width=3.0cm,minimum height=1.1cm,fill=black!3]
 (others) at (14.6,-2.2) {Other blockchain\\instances};
\draw[branch] (service.south) -- (8.5,-0.75) -- (6.3,-0.75) -- (chain.north);
\draw[branch] (8.5,-0.75) -- (14.6,-0.75) -- (others.north);
\node[role] (geth) at (1.9,-4.95) {\textbf{Execution nodes}\\Geth};
\node[role] (beacon) at (6.3,-4.95) {\textbf{Beacon nodes}\\Lighthouse};
\node[role] (validator) at (10.7,-4.95) {\textbf{Validator clients}\\Lighthouse};
\node[role,fill=orange!7] (faucet) at (15.1,-4.95) {\textbf{Faucet}};
\draw[branch] (chain.south) -- (6.3,-3.95);
\draw[branch] (1.9,-3.95) -- (15.1,-3.95);
\foreach \name in {geth,beacon,validator,faucet}{
 \draw[branch] (\name.north) -- (\name.north |- 1.9,-3.95);
}
\end{tikzpicture}}
\caption{Chain-wide configuration and logical participants within the Ethereum service.}
\Description{A tree starts with Ethereum service and branches to a blockchain instance and other chain instances. Four separate ellipses inside the blockchain instance show Protocol settings, Genesis state, Accounts, and Initial validators. The first instance expands into execution nodes using Geth, beacon nodes using Lighthouse, validator clients using Lighthouse, and a faucet. The branches represent configuration ownership and service organization rather than software inheritance or runtime API calls.}
\label{fig:ethereum-service-architecture}
\end{figure*}
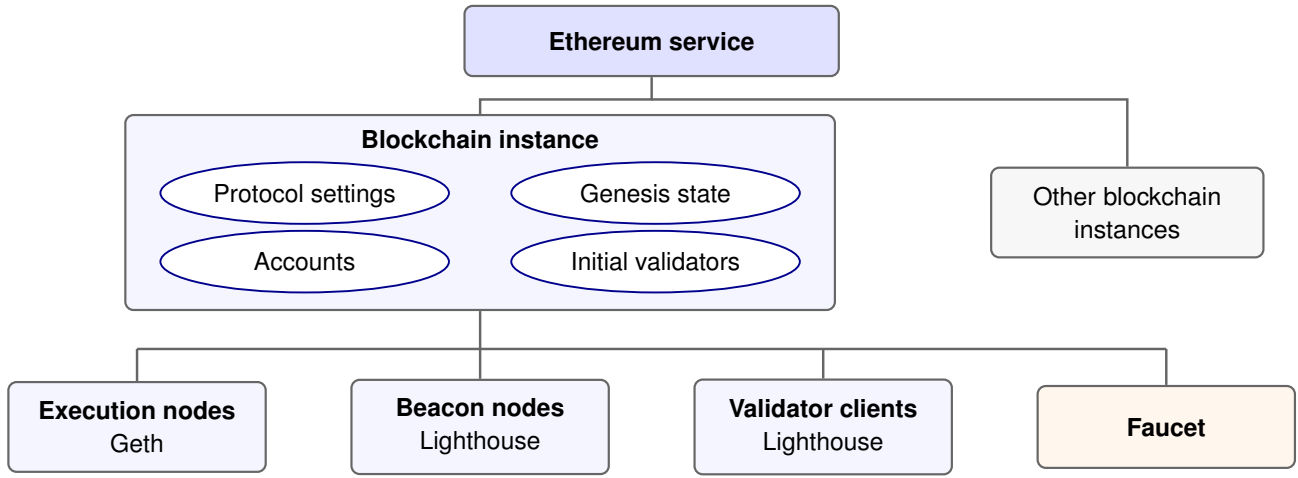

Ethereum currently uses the PoS (Proof of Stake) consensus protocol, and we focus on this
configuration when describing our Ethereum service. The service also supports PoW (Proof of
Work) and PoA (Proof of Authority) used in the earlier Ethereum
blockchain~\cite{ethereum-merge,ethereum-poa}. Ethereum service organizes the system at two
levels: blockchain and the participating nodes. As illustrated in
Figure~\ref{fig:ethereum-service-architecture}, the blockchain component holds the common
configuration of the entire blockchain, organizes all types of nodes (execution nodes, beacon
nodes, and validator nodes), and establishes their relationships. Node-specific roles, endpoints, and client options belong to individual participants. This organization lets the service coordinate the chain's initial state and client connections while preserving independent control over each participant's deployment and resource settings.

Within each chain, execution, beacon, and validator clients are defined as separate
participants, and the service records their associations through virtual-node names.
Listing~\ref{lst:ethereum-service} shows how a beacon node is linked to an execution client and
a validator client to a beacon node without specifying their deployment hosts. Our reference
deployment uses Geth and Lighthouse, with a dedicated execution client for each beacon node
(following Lighthouse's one-to-one connection requirement~\cite{lighthouse-node-deployment}). The service coordinates shared execution and consensus genesis states, validator initialization materials, and startup dependencies so that these separately configured clients begin operating on the same chain.

\noindent\begin{minipage}[t]{\linewidth}
\begin{lstlisting}[style=chainseed,
  caption={Client roles and their logical connections within the chain.},
  label={lst:ethereum-service}]
eth = EthereumService()
chain = eth.createBlockchain("experiment", ConsensusMechanism.POS)
gethServer = chain.createGethNode("gethServer").setBootNode(True)
beaconServer = chain.createBeaconNode("beaconServer").setBootNode(True)
beaconServer.connectToGethNode("gethServer")
validatorServer = chain.createVcNode("validatorServer")
validatorServer.connectToBeaconNode("beaconServer")
\end{lstlisting}
\end{minipage}
\par

\subsubsection{Extending the Service Design}
\label{sec:chain-extension}
Following the Ethereum service's organization of chain configuration and participating nodes, we extended ChainSEED to Monero and Solana. Their services adapt client configuration and initialization to each protocol while reusing SEED's virtual-node bindings and compilation pipeline. We implemented both services within two weeks. As illustrated in Figure~\ref{fig:chainseed-software-architecture}, their nodes can be deployed alongside Ethereum clients in the same SEED network. Section~\ref{sec:extension-evaluation} evaluates deployments using these services.

\subsection{Customization}
\label{sec:customization}
Blockchain studies may require a variety of customization, such as predefined account balances,
initial validator populations, selected peer connections, or modified client software. ChainSEED supports these customizations in its design.

\subsubsection{Customization of the Genesis Configuration}
\label{sec:genesis-customization}
\label{sec:account-balance}
The genesis configuration determines the blockchain's starting state. Ethereum has two genesis
states: the execution genesis and the consensus genesis. ChainSEED supports the customization of
both genesis states, including the chain ID, initial block gas limit, account balances, and the
initial validator population. Users can supply an execution genesis configuration and add
account allocations through the provided APIs. They can also add a list of validators to the
consensus genesis. Listing~\ref{lst:genesis-customization} illustrates these declarations using the chain and validator client defined in Listing~\ref{lst:ethereum-service}. The service uses these declarations to prepare the shared execution and consensus genesis states.

\begin{lstlisting}[style=chainseed,
  caption={Customizing execution genesis and the initial validator population.},
  label={lst:genesis-customization}]
chain.setGenesis(genesis_json)
chain.addLocalAccount("Account Address", balance=100)
chain.createBeaconSetupNode("setupServer")
validatorServer.enablePOSValidatorAtGenesis()
\end{lstlisting}

ChainSEED also provides a faucet for funding accounts after startup. The faucet transfers funds
through an associated execution client's RPC interface, allowing wallet programs and scripts to obtain funds without regenerating the chain's genesis. Listing~\ref{lst:faucet-funding}
illustrates how the faucet is added to the emulator.

\begin{lstlisting}[style=chainseed,
  caption={Adding a faucet for account funding on the running chain.},
  label={lst:faucet-funding}]
chain.createFaucetServer("faucetServer", ...)
\end{lstlisting}

\subsubsection{Customization APIs}
\label{sec:customization-api}
ChainSEED also supports several other types of customization, including peer configuration and
user-supplied client implementations. Protocol-level peer options and software-level client settings or modified binaries provide
parallel inputs to deployment. The service combines them into configuration files and startup
commands for the selected software on bound SEED hosts. Section~\ref{sec:applications} will present studies enabled by these capabilities.

Blockchain studies may need to vary peer discovery and connectivity. ChainSEED lets users
designate bootstrap nodes, set peer limits, and even specify selected P2P connections. 
Moreover, some experiments require changes to the blockchain software's behavior. Users can
adjust the existing software options or supply a modified binary to replace the existing
software implementation. The service will prepare the software's configuration and startup commands, allowing customized clients to participate in the blockchain.

\subsection{Experimental Interaction Tools}
\label{sec:tools}
Researchers need to generate experimental workloads, interact with blockchain clients, and observe how
the chain and its nodes respond. ChainSEED provides these functionalities through a monitoring
system, a configurable transaction generator, and the ability to integrate with the existing blockchain tools.

\subsubsection{Monitoring System}
\label{sec:monitoring-system}
\label{sec:workloads-observation}
\label{sec:visualization}

Our monitoring system includes EthExplorer and Ethereum Lookingglass. EthExplorer is based on
the open-source beaconcha.in explorer~\cite{beaconchain-explorer}. It visualizes blockchain
data, including blocks, transactions, account activities, and validator proposals and
attestations (Appendix~\ref{sec:ethexplorer-views}). For the node-level visualization, we 
developed Ethereum Lookingglass. Users can use this tool to visualize peer connections and client dependencies; 
they can also inspect each node's resource usage, beacon processor queues, and Geth transaction pools. 
Recorded block and transaction events can also be replayed alongside these node observations. 
Appendix~\ref{sec:lookingglass-views} shows a view of the Lookingglass.

\subsubsection{Transaction Generation}
\label{sec:transaction-generation}
To help researchers generate experimental workloads on the blockchain, we have developed a
transaction generation tool. Ethereum transactions range from simple value transfers to smart-contract interactions with different execution demands. To reflect this diversity, our transaction generator submits transfers, contract calls, and contract deployments through client RPC interfaces. Its default sampling weights, listed in Appendix~\ref{sec:transaction-mix-appendix}, are derived from successful Ethereum mainnet transactions in Xatu's public execution-layer dataset~\cite{xatu-execution-data}. Users can adjust the transaction mix and submission rate independently. All Ethereum transaction-load experiments in Sections~\ref{sec:scalability} and~\ref{sec:evaluation} use the default mixture.
\label{par:transaction-generator}

\subsubsection{Integration with Existing Software}
\label{sec:software-integration}
Researchers may also need to develop and test blockchain applications using familiar software. ChainSEED supports these workflows through the standard interfaces exposed by its real Ethereum clients. Users can connect MetaMask to the execution client's JSON-RPC endpoint for account queries and transaction submission~\cite{geth-rpc,metamask-local-node}. Remix uses this wallet connection to deploy and interact with smart contracts~\cite{remix-deploy-run}, while Hardhat connects to the RPC endpoint for scripted deployment and testing. These integrations support both interactive application use and automated workflows within the emulated environment. Appendix~\ref{sec:software-fidelity} presents examples of the wallet, contract-interaction, and scripted-testing workflows.
\paragraph{AI Usage Disclosure.}
We used generative AI to revise portions of the manuscript and improve grammar and spelling.

\section{Scalability Analysis and Optimization}
\label{sec:scalability}
Ethereum brings together thousands of execution and consensus nodes, approximately 900,000 active validator identities, and sustained transaction activity at around 20~transactions per second (TPS)~\cite{ethernodes-scale,rated-scale,chainspect-scale}. Studies of its operation may therefore require a sufficiently large node network to examine consensus behavior~\cite{available-attestation-2025}, transaction workloads that exercise execution capacity~\cite{mhot-2026}, or a large validator population to investigate participation and deployment~\cite{deanonymizing-validators-2025}. Given our design of running real blockchain software, hardware resources become a primary constraint on meeting these experimental demands. Motivated by these scalability challenges and researchers' needs, we formulate three research questions:
\begin{enumerate}
\item \textbf{RQ1:} How can we support more consensus nodes within a given memory budget while maintaining timely consensus processing and chain synchronization?
\item \textbf{RQ2:} How can execution clients sustain higher transaction throughput while limiting transaction backlog and maintaining stable chain operation?
\item \textbf{RQ3:} How can we support more validator identities with lower resource overhead while maintaining timely duty execution and healthy consensus?
\end{enumerate}

All experiments in this section run on a server with one AMD EPYC 9654 processor (96 physical cores, 192 logical threads) and approximately 600~GiB of memory.

\subsection{Reducing Beacon Node Memory}
To address \textbf{RQ1}, we first examine how to reduce each consensus client's memory cost to accommodate a larger consensus network on the same hardware. Every additional node brings its own memory footprint, so even modest savings per node can create room for a larger deployment. The question is which parts of this footprint can be reduced while preserving the node's ability to keep pace with the chain. We begin with the configuration options exposed by Lighthouse, investigating whether its existing controls can reduce the memory cost of running each node.

To identify which configurations best reduce beacon-node memory usage and thereby support a larger node network, we compare Lighthouse command-line configurations governing data retention, network connections, service functionality, and task processing. Using the unoptimized configuration as a control, we examine the effect of each type of adjustment on per-node memory. Each experiment deploys 300 beacon nodes, 300 execution nodes, and 900 validator clients under a 1~TPS workload, and measures per-node memory over ten hours. Table~\ref{tab:memory-configurations} summarizes changes in mean and P95 memory relative to the baseline.


\begin{table}[!htb]
\centering\small
\begin{tabularx}{\linewidth}{@{}>{\raggedright\arraybackslash}Xrr@{}}
\toprule
Adjustment & \shortstack{Mean memory\\change} & \shortstack{P95 memory\\change}\\
\midrule
Baseline & $0\%$ & $0\%$\\
Cache limits & $-2.1\%$ & $-1.1\%$\\
Peer limits & $-2.4\%$ & $-0.1\%$\\
Observer-only historical-state reconstruction & {\boldmath$-12.0\%$} & {\boldmath$-13.1\%$}\\
Internal cache sizing & $+0.2\%$ & $-1.0\%$\\
Disable light-client support & $-0.2\%$ & $+0.5\%$\\
Parallel-processing threads (Rayon) & $-1.5\%$ & $-1.8\%$\\
Disable QUIC transport & $+0.1\%$ & $-0.5\%$\\
Worker cap = 2 & {\boldmath$-41.5\%$} & {\boldmath$-54.6\%$}\\
Processing-queue limits & $-1.2\%$ & $0.0\%$\\
Gossip-message batching & $+1.2\%$ & $+0.8\%$\\
\bottomrule
\end{tabularx}
\caption{Changes in beacon-node memory relative to the baseline.}
\floatnote{P95 is the value at or below which 95\% of measurements fall.}
\label{tab:memory-configurations}
\end{table}

Table~\ref{tab:memory-configurations} identifies two adjustments that substantially reduce beacon-node memory: restricting historical-state reconstruction to observer nodes and limiting beacon-processor worker concurrency. In contrast, all other tested adjustments change mean memory by at most 2.4\%. We examine the memory benefits and operational implications of these two adjustments below.

Retaining historical-state reconstruction only at the observer reduces mean per-node memory by 12.0\%. Ordinary nodes do not need to reconstruct past states for historical queries in order to follow and validate the current chain. We can therefore allocate this capability according to each node's role: one or two designated nodes serve the historical queries required by our blockchain visualization tool, EthExplorer, while the remaining nodes avoid this memory cost, leaving room for a larger consensus network. Experiments that study historical-state APIs may instead require reconstruction at multiple nodes, for example to compare query behavior under different network conditions and query loads. In these cases, researchers can retain the capability at the nodes under study and allocate resources according to the historical-query coverage and deployment scale their experiments require.

The worker cap limits how many workers can process tasks concurrently in Lighthouse's Beacon Processor. Reducing it from the default setting to two lowers mean memory by 41.5\% and P95 memory by 54.6\%, to 0.48 and 0.54~GiB per beacon node, respectively. To understand this substantial reduction, we examined how Lighthouse determines worker concurrency. Its default worker limit depends on the number of logical CPUs visible to the process. In our deployment, each beacon-node container sees all 192 logical CPUs of the host, so the default concurrency reflects the CPU count of the shared host rather than resources dedicated to an individual node. However, these memory savings may come at the cost of processing capacity. We therefore compare different worker caps to examine the trade-off between memory savings and chain-synchronization performance.

To explore this trade-off, we vary the worker cap across 1, 2, 4, 8, and 16 while holding the remaining settings fixed. We retain the deployment scale and 1~TPS workload described above and run each configuration for ten hours. Alongside per-node memory, we track differences in chain progress across 32 beacon nodes, using the gap between the highest and lowest head slots reported in each observation round. This comparison reveals whether further memory savings are accompanied by a larger divergence in node progress.

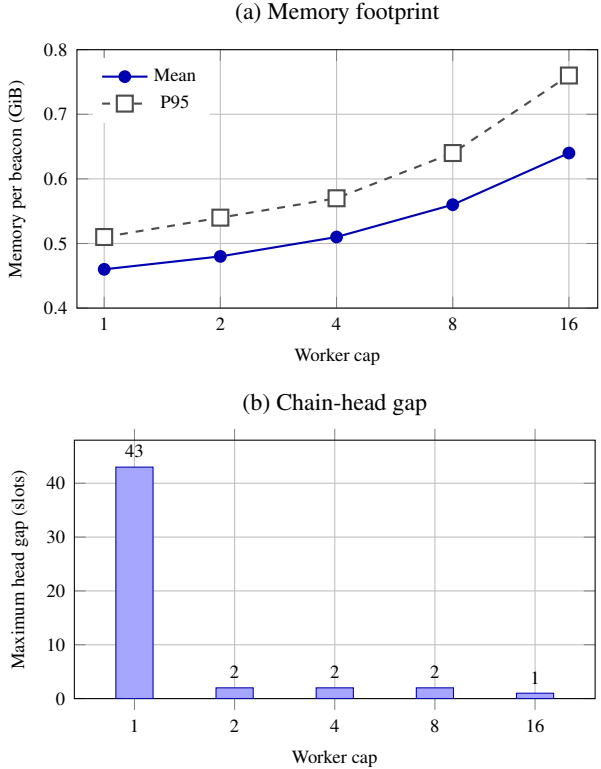
\begin{figure}[!htb]
\centering
\begin{minipage}[t]{\linewidth}
\centering
\begin{tikzpicture}
\begin{axis}[
 width=\linewidth,height=5cm,
 xmode=log,log basis x=2,xtick={1,2,4,8,16},
 xticklabels={1,2,4,8,16},xmin=0.85,xmax=19,
 ymin=0.4,ymax=0.8,
 xlabel={Worker cap},ylabel={Memory per beacon (GiB)},
 title={(a) Memory footprint},
 tick label style={font=\scriptsize},label style={font=\scriptsize},
 title style={font=\small},grid=major,
 legend style={font=\scriptsize,draw=none,at={(0.03,0.97)},anchor=north west}]
\addplot[blue!70!black,mark=*,thick] table[x=cap,y=memory_mean_gib,col sep=comma]{data/worker-cap-low-load.csv};
\addlegendentry{Mean}
\addplot[black!70,mark=square*,mark size=3pt,mark options={solid,fill=white,line width=0.8pt},dashed,thick] table[x=cap,y=memory_p95_gib,col sep=comma]{data/worker-cap-low-load.csv};
\addlegendentry{P95}
\end{axis}
\end{tikzpicture}
\end{minipage}\par\smallskip
\begin{minipage}[t]{\linewidth}
\centering
\begin{tikzpicture}
\begin{axis}[
 width=\linewidth,height=5cm,
 symbolic x coords={1,2,4,8,16},xtick=data,enlarge x limits=0.15,
 ybar,bar width=14pt,
 ymin=0,ymax=48,ytick={0,10,20,30,40},
 xlabel={Worker cap},ylabel={Maximum head gap (slots)},
 title={(b) Chain-head gap},
 tick label style={font=\scriptsize},label style={font=\scriptsize},
 title style={font=\small},grid=major]
\addplot[draw=blue!70!black,fill=blue!35,
 nodes near coords,point meta=y,every node near coord/.append style={font=\scriptsize}]
 table[x=cap,y=max_head_spread_slots,col sep=comma]{data/worker-cap-low-load.csv};
\end{axis}
\end{tikzpicture}
\end{minipage}
\caption{Beacon-node memory and maximum chain-head gap across worker caps.}
\floatnote{P95 is the value at or below which 95\% of memory measurements fall.}
\Description{Two panels show mean and P95 memory per beacon node and the maximum observed difference in reported chain-head slots as the worker cap changes.}
\label{fig:rq-memory}
\end{figure}

Figure~\ref{fig:rq-memory} shows that further reducing the worker cap does not necessarily yield a favorable trade-off. Lowering the cap from two to one saves only another 0.02~GiB per beacon node, but the observed chain-head gap reaches 43 slots and exceeds eight slots in 2.06\% of valid observation rounds. In contrast, the maximum gap remains within two slots for all tested caps of two or higher, while memory consumption increases with the cap. In this low-load experiment, cap=2 therefore combines a low memory footprint with the absence of the pronounced differences in node progress observed at cap=1. We use cap=2 as the starting point for further investigation under higher transaction loads.

The low-load results make cap=2 a promising option for reducing beacon-node memory, but leave open whether its limited concurrency can accommodate a busier chain. We therefore compare caps 2, 4, 8, and 16 under a changing transaction load. We examine whether high-priority API work accumulates and whether the chain continues to process transactions and advance consensus. Each configuration runs separately at the same deployment scale. The workload starts at 15~TPS for four hours, rises to 20~TPS for two hours, and returns to 15~TPS for four hours, allowing us to observe both accumulation and recovery in this queue.

\begin{table*}[!tb]
\centering\small
\setlength{\tabcolsep}{4pt}
\begin{tabular}{lrrrrrrrr}
\toprule
 & \multicolumn{4}{c}{Mean waiting tasks} & \multicolumn{4}{c}{P95 bucket upper bound}\\
\cmidrule(lr){2-5}\cmidrule(lr){6-9}
Phase & Cap 2 & Cap 4 & Cap 8 & Cap 16 & Cap 2 & Cap 4 & Cap 8 & Cap 16\\
\midrule
Initial load & 1.302 & 0.500 & 0.012 & 0.002 & 16 & 4 & 0 & 0\\
Higher load & 1.409 & 0.573 & 0.017 & 0.001 & 16 & 4 & 0 & 0\\
Recovery & 1.252 & 0.474 & 0.011 & 0.002 & 16 & 4 & 0 & 0\\
\bottomrule
\end{tabular}
\caption{Queue length for high-priority API work across workload phases.}
\floatnote{API P0 is the Beacon Processor's high-priority API work queue. Means average observed queue lengths. P95 denotes the 95th percentile, reported here as its histogram bucket upper bound in tasks.}
\label{tab:worker-queue}
\end{table*}

As shown in Table~\ref{tab:worker-queue}, smaller worker caps leave more high-priority API work waiting to be processed. At cap=2, the mean observed queue length rises from 1.302 to 1.409 tasks as the load increases, then falls to 1.252 when the load returns to 15~TPS. Cap=4 follows the same pattern with shorter queues, while the means at caps 8 and 16 remain close to zero. The lower-cap configurations therefore incur more queueing, but the increase during the higher-load phase subsides as the workload falls.

In return, cap=2 uses 0.494~GiB per beacon node on average, compared with 0.660~GiB at cap=16 (Table~\ref{tab:worker-chain}). Despite the additional queueing, all four runs maintain similar confirmation times. All submitted transactions are eventually included and executed successfully, and finalization continues. Their maximum observed chain-head gaps remain within two slots. The cap=2 run exhibits occasional missed slots and slightly lower validator activity, but the chain continues to operate under the tested workload. Lower worker caps therefore offer memory savings at the cost of additional waiting in the high-priority API queue. Users can choose a smaller cap to conserve memory or increase it when limiting such queueing is a priority.

\begin{table*}[!tb]
\centering\small
\setlength{\tabcolsep}{3pt}
\begin{tabularx}{\linewidth}{@{}r*{7}{>{\centering\arraybackslash}X}@{}}
\toprule
Cap & \multicolumn{2}{c}{Beacon memory (GiB)} & Block visibility & Active validators & Maximum head gap & Missed slots & Confirmation\\
\cmidrule(lr){2-3}
 & Mean & P95 & P95 (s) & Mean (\%) & (slots) & (\%) & P95 (s)\\
\midrule
2 & 0.494 & 0.556 & 1.825 & 99.953 & 2 & 0.101 & 16.012\\
4 & 0.520 & 0.583 & 1.809 & 100.000 & 1 & 0.000 & 16.023\\
8 & 0.575 & 0.653 & 1.820 & 100.000 & 1 & 0.000 & 16.001\\
16 & 0.660 & 0.782 & 2.049 & 100.000 & 1 & 0.000 & 16.112\\
\bottomrule
\end{tabularx}
\caption{Beacon-node memory and chain operation across worker caps.}
\floatnote{Validator activity is the fraction of validator identities reported as active. Missed slots have no observed block. Block visibility measures the time from the first block observation until 95\% of sampled beacon nodes report the same block. Confirmation measures the time from RPC acceptance to receipt observation. P95 is the value at or below which 95\% of the corresponding measurements fall.}
\label{tab:worker-chain}
\end{table*}


\subsection{Improving Execution-Client Transaction Throughput}
Beyond expanding the node network, \textbf{RQ2} asks how execution clients can sustain transaction loads approaching real-world operation. Because transactions are processed by real execution clients, increasing the generator's sending rate does not ensure that the chain can process transactions at the same rate. A sustained excess of arrivals over processing capacity leads to a backlog. Scaling transaction load therefore depends on the on-chain throughput that execution clients can sustain. Geth continually allocates and reclaims memory during transaction execution and state maintenance. When the runtime memory budget is too tight, garbage collection may run more frequently, consuming CPU time that could otherwise serve transaction processing. Relaxing the budget may reduce this overhead, but can also increase a node's memory footprint~\cite{go-gc-guide}. We therefore vary \texttt{GOMEMLIMIT}, the Go runtime's soft memory limit, to examine the transaction loads supported by different per-node memory configurations, together with their memory costs and chain performance, and identify suitable resource configurations for different workload requirements.

We first compare memory configurations under the same transaction load to examine their effects on transaction processing and resource consumption. Using the deployment scale described above, we set \texttt{GOMEMLIMIT} to 256, 384, and 512~MiB while keeping the beacon worker cap at two and the execution-client cache at 128~MiB. Each configuration runs independently, with a one-hour warm-up followed by eight hours at 15~TPS and a one-hour recovery phase at 1~TPS.

\begin{table*}[!tb]
\centering\small
\setlength{\tabcolsep}{3pt}
\begin{tabularx}{\linewidth}{@{}r*{6}{>{\centering\arraybackslash}X}@{}}
\toprule
\texttt{GOMEMLIMIT} & \multicolumn{2}{c}{Execution memory (GiB)} & Included throughput & Confirmation & Block visibility & Missed blocks\\
\cmidrule(lr){2-3}
(MiB) & Mean & P95 & (TPS) & P95 (s) & P95 (s) & (\%)\\
\midrule
256 & 0.500 & 0.638 & 14.970 & 29.732 & 7.267 & 2.583\\
384 & 0.588 & 0.702 & 14.993 & 16.409 & 2.850 & 0.042\\
512 & 0.635 & 0.753 & 15.000 & 16.358 & 2.722 & 0.000\\
\bottomrule
\end{tabularx}
\caption{Execution-client memory, throughput, and missed blocks.}
\floatnote{Memory is the container's cgroup usage, whereas \texttt{GOMEMLIMIT} applies to the Go runtime. Included throughput counts transactions incorporated into blocks per second during the load window. Confirmation measures the time from RPC acceptance to receipt observation. P95 is the value at or below which 95\% of measurements fall.}
\label{tab:geth-memory}
\end{table*}

\begin{table*}[!tb]
\centering\small
\begin{tabular}{rrrrrr}
\toprule
\texttt{GOMEMLIMIT} & Target & Sent & Included & Final inclusion & Confirmation\\
(MiB) & (TPS) & (TPS) & (TPS) & (\%) & P95 (s)\\
\midrule
384 & 20 & 20.000 & 19.999 & 100.000 & 16.655\\
384 & 25 & 25.000 & 24.998 & 100.000 & 17.096\\
\midrule
512 & 20 & 20.000 & 20.000 & 100.000 & 16.589\\
512 & 25 & 25.000 & 24.998 & 100.000 & 16.253\\
\bottomrule
\end{tabular}
\caption{Transaction throughput and inclusion at higher offered loads.}
\floatnote{Included TPS covers the load window. Final inclusion is the fraction of submitted transactions eventually incorporated into the chain, including after the load ends. Confirmation measures the time from RPC acceptance to receipt observation. P95 is the value at or below which 95\% of confirmation measurements fall.}
\label{tab:geth-load}
\end{table*}

Table~\ref{tab:geth-memory} shows that although all three configurations eventually include every transaction sent during the load phase, the tighter memory budget is accompanied by longer confirmation times and more missed blocks. Raising \texttt{GOMEMLIMIT} from 256 to 384~MiB increases mean memory by only 0.088~GiB per execution node. Confirmation P95 falls from 29.732 to 16.409~s, the missed-block rate drops from 2.583\% to 0.042\%, and included throughput moves closer to the 15~TPS input. Increasing the limit further to 512~MiB consumes additional memory and eliminates missed blocks in this run, while confirmation P95 remains close to that at 384~MiB. The 384~MiB configuration thus offers a smaller footprint than 512~MiB while keeping pace with the offered transaction load, making it a candidate for further exploration at higher loads.

The near-15~TPS throughput at 384~MiB leaves open how much additional transaction load this memory budget can accommodate. We next increase the input to 20 and 25~TPS to examine whether the execution clients continue to keep pace and whether a larger memory budget also supports these higher loads.

As shown in Table~\ref{tab:geth-load}, the 384~MiB configuration continues to process incoming transactions when the load rises to 25~TPS. Included throughput reaches 24.998~TPS over eight hours, all load-phase transactions are eventually included, and confirmation P95 is 17.096~s. Thus, besides improving chain operation at 15~TPS, the 384~MiB configuration also supports the higher load of 25~TPS.

At 512~MiB, included throughput likewise reaches 24.998~TPS, all submitted transactions are eventually included, and confirmation P95 is 16.253~s. Further increasing the sustained transaction load requires more transactions per block, raising gas usage and potentially exceeding the EIP-1559 target. Persistent usage above this target raises the base fee, potentially pushing transaction costs beyond senders' budgets even when runtime memory is sufficient~\cite{eip-1559}. ChainSEED therefore allows users to raise the block gas limit through its customization API, expanding capacity and easing fee pressure at the same load.

\subsection{Multiplexing Validators}
Validators propose blocks and submit attestations that contribute to chain selection and finality in Ethereum PoS~\cite{ethereum-pos-background}. The public network contains approximately 900,000 active validator identities~\cite{rated-scale}. Studying their participation, duty distribution, and deployment therefore requires emulated populations large enough to represent these behaviors~\cite{deanonymizing-validators-2025}. \textbf{RQ3} asks how to support these populations with lower resource overhead while preserving timely duty execution and consensus progress.

\begin{figure}[!htb]
\centering
\input{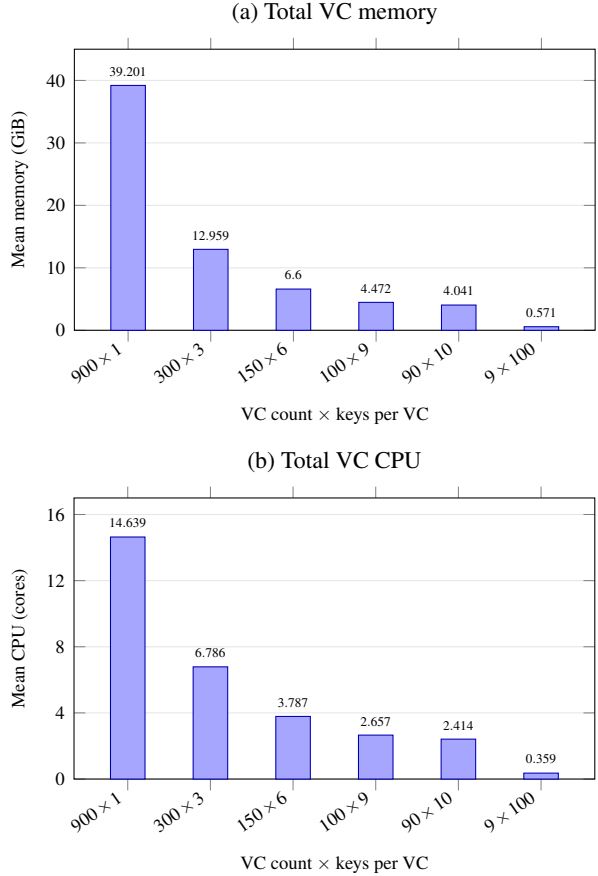}
\caption{Aggregate validator-client memory and CPU usage under multiplexing.}
\floatnote{CPU usage is expressed in logical-core equivalents, with one core representing one fully utilized logical CPU.}
\Description{Two bar charts show mean aggregate validator-client memory and CPU usage as more validator keys share each client.}
\label{fig:rq-validators}
\end{figure}

Emulating a larger validator population introduces another source of resource demand. If each validator identity is assigned a separate client, expanding the population also increases the number of client processes, each with its own memory and CPU overhead. Yet a validator identity is defined by its key and consensus duties, rather than by a dedicated process. Lighthouse already supports assigning multiple validator keys to one client while preserving their individual identities and duties~\cite{lighthouse-validator-management}. We use this configuration for validator multiplexing and examine whether hosting the same validator population in fewer clients can reduce resource consumption without impairing consensus operation.

Using a deployment similar to the preceding experiments, we keep 900 validator identities and vary how their keys are distributed across validator clients (VCs). We compare six configurations, expressed as VC count $\times$ keys per VC: $900\times1$, $300\times3$, $150\times6$, $100\times9$, $90\times10$, and $9\times100$. This progressively consolidates the same validator population into fewer clients while retaining all 300 beacon nodes and 300 execution nodes. Under a 1~TPS mixed-transaction workload, we compare aggregate VC resource usage and consensus progress over a five-hour observation window.

As shown in Figure~\ref{fig:rq-validators}, increasing the number of keys managed by each client progressively reduces aggregate VC memory and CPU usage. From $900\times1$ to $9\times100$, mean memory falls from 39.201 to 0.571~GiB and mean CPU usage from 14.639 to 0.359 cores, reductions of 98.54\% and approximately 97.55\%, respectively. With multiplexing, multiple validator identities share a client process, its runtime, and its connections to beacon nodes instead of maintaining a separate client for each identity. This sharing reduces duplicated client overhead while preserving the validator population, allowing the same 900 identities to be hosted at a much lower resource cost.

These resource savings raise a further question: can fewer clients still maintain validator participation and consensus progress? Across all six configurations, all 900 validator identities remain active in the recorded observations, and P95 attestation inclusion delay remains one slot. As shown in Table~\ref{tab:validator-multiplex}, P95 finality lag remains two epochs. Its maximum rises from two to three epochs in the four more consolidated configurations, but finalization continues to advance in every run. The maximum chain-head gap remains one slot across sampled beacon nodes. All 18,000 submitted transactions are eventually included in each configuration. Together, these observations show that the substantial reduction in VC resource usage is not accompanied by evident deterioration in validator participation or consensus progress under the tested workload.

\begin{table*}[!tb]
\centering\small
\setlength{\tabcolsep}{3pt}
\begin{tabularx}{\linewidth}{@{}r*{5}{>{\centering\arraybackslash}X}@{}}
\toprule
VCs $\times$ keys per VC & Active validators & Attestation inclusion & Finality lag & Block-observation spread & Maximum head gap\\
 & (\%) & P95 (slots) & P95 / max (epochs) & P95 (s) & (slots)\\
\midrule
$900\times1$ & 100 & 1 & 2 / 2 & 2.630 & 1\\
$300\times3$ & 100 & 1 & 2 / 2 & 2.747 & 1\\
$150\times6$ & 100 & 1 & 2 / 3 & 2.706 & 1\\
$100\times9$ & 100 & 1 & 2 / 3 & 2.711 & 1\\
$90\times10$ & 100 & 1 & 2 / 3 & 2.718 & 1\\
$9\times100$ & 100 & 1 & 2 / 3 & 2.678 & 1\\
\bottomrule
\end{tabularx}
\caption{Validator participation and consensus progress under client multiplexing.}
\floatnote{Validator activity is the percentage of validator identities observed as active. Attestation inclusion delay counts slots from an attestation's assigned slot to its inclusion in a block. Finality lag is the gap between the head and finalized epochs. Block-observation spread is the difference between the earliest and latest observations of the same block across sampled beacon nodes. P95 is the value at or below which 95\% of the corresponding measurements fall.}
\label{tab:validator-multiplex}
\end{table*}

Validator multiplexing provides a practical way to reduce the resource cost of hosting a large validator population. Assigning 100 keys to each VC substantially reduces client resource usage while maintaining normal consensus progress. However, these validators also share the fate of their client: if it fails, all validator identities it manages may be unable to perform their duties. Users should therefore weigh the resource savings against the number of validators affected by a single client failure when choosing the multiplexing degree.

\section{Evaluation}
\label{sec:evaluation}
Section~\ref{sec:scalability} examines how client configurations affect resource use and blockchain operation. We combine selected configurations to evaluate larger Ethereum deployments on a single server, measuring resource requirements, transaction processing, and consensus progress. We also evaluate Monero and Solana deployments to assess how the shared service design supports experiments beyond Ethereum.

\subsection{Scalability: Large-Scale Deployment on a Single Server}
\label{sec:combined-scale}
To plan larger Ethereum experiments, researchers need to understand the resource cost of hosting both the node network and the validator population. We therefore increase these populations together on a single server under a sustained transaction load. The server has two AMD EPYC 9654 processors (192 physical cores, 384 logical threads) and approximately 1,000~GiB of memory. By measuring memory and CPU consumption alongside transaction processing and consensus progress, we provide a practical reference for choosing deployment sizes within a hardware budget.

We use a worker cap of 2 for beacon nodes, a \texttt{GOMEMLIMIT} of 384~MiB for execution clients, and 100 validator identities per validator client. Each beacon node is paired with a dedicated execution node, and the numbers of beacon nodes, execution nodes, and validator clients follow a $1:1:1.5$ ratio. The four deployments contain 100, 200, 300, and 400 beacon--execution pairs, corresponding to 15,000, 30,000, 45,000, and 60,000 validator identities, respectively. Each deployment is observed for five hours under an aggregate transaction input of 20~TPS.

\begin{figure*}[!tb]
\centering
\includegraphics[width=\linewidth]{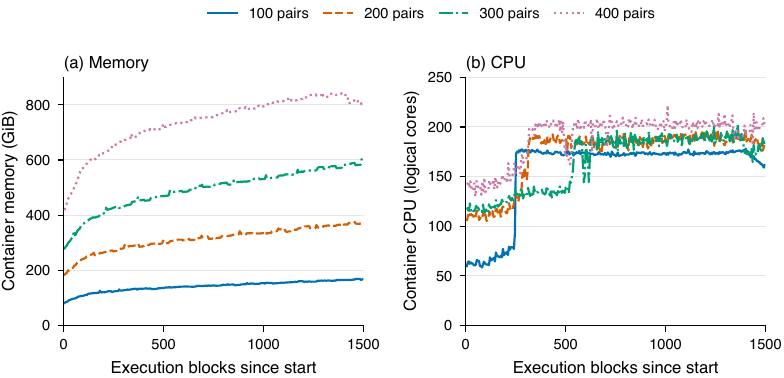}
\caption{Memory and CPU usage during large-scale blockchain deployments.}
\Description{Two line plots show container memory and CPU usage against chain progress, with separate curves for the four deployment sizes.}
\label{fig:combined-scale-resources}
\end{figure*}

Figure~\ref{fig:combined-scale-resources} shows that memory demand increases with deployment size and grows during the observation windows. Mean container memory rises from 141.9~GiB at 100 beacon--execution pairs to 313.7, 493.1, and 738.4~GiB at 200, 300, and 400 pairs, respectively. Mean CPU usage is 155.7, 172.8, 164.8, and 189.4 logical cores across the four deployments, respectively. The 400-pair deployment reaches a peak container memory usage of 844.7~GiB.

As the node network and validator population grow, we examine whether the chain continues to process its transaction workload and advance consensus. We track transaction inclusion, validator activity, missed slots, and finalization progress during the same runs. Table~\ref{tab:configuration-boundaries} summarizes these measurements across deployment sizes.

\begin{table*}[!tb]
\centering\small
\begin{tabularx}{\linewidth}{@{}r*{4}{>{\centering\arraybackslash}X}@{}}
\toprule
Beacon--execution pairs & Included TPS & Mean active validators (\%) & Finalized epoch advance & Missed slots (\%)\\
\midrule
100 & 19.990 & 99.475 & 47 & 0.000\\
200 & 19.988 & 97.069 & 47 & 0.000\\
300 & 19.984 & 99.577 & 47 & 0.000\\
400 & 19.973 & 99.095 & 47 & 0.200\\
\bottomrule
\end{tabularx}
\caption{Transaction processing and consensus progress across deployment sizes.}
\floatnote{Validator activity is the fraction of validator identities observed as active at the reference beacon node. Finalized epoch advance is the increase in the finalized epoch over the observation window.}
\label{tab:configuration-boundaries}
\end{table*}

On the tested server, ChainSEED supports an Ethereum deployment of 400 beacon nodes, 400 execution nodes, and 60,000 validator identities under a 20~TPS input. This expanded deployment continues to process transactions and advance consensus: across all four deployment sizes, transaction inclusion remains close to the input rate, and finalized epochs advance by 47 during each five-hour observation window. All submitted transactions eventually execute successfully. At the largest scale, mean validator activity is 99.1\%, and the missed-slot rate is 0.2\%, providing further evidence of continued validator participation and block production.


\subsection{Scalability Beyond Ethereum}
\label{sec:extension-evaluation}

\paragraph{Monero.}
Proof-of-work experiments bring an additional resource demand to large node networks: participating miners must perform the computation needed to produce blocks. Using Monero, we evaluate how ChainSEED supports this activity alongside a 1,200-node network that validates and propagates transactions and blocks. We vary the number of active miners to examine resource requirements and whether the network continues to process transactions and advance the chain.

\begin{table}[!htb]
\centering\small
\begin{tabularx}{\linewidth}{@{}r*{4}{>{\centering\arraybackslash}X}@{}}
\toprule
Target miners & Mean memory (GiB) & Mean CPU (logical cores) & Height advance & Included transfers (\%)\\
\midrule
1 & 87.829 & 8.858 & 287 & 100.0\\
10 & 89.696 & 18.565 & 753 & 100.0\\
100 & 96.390 & 102.830 & 394 & 100.0\\
200 & 100.372 & 186.812 & 324 & 100.0\\
\bottomrule
\end{tabularx}
\caption{Resource consumption and chain progress in the 1,200-node Monero network.}
\floatnote{Resource totals include Monero and network-infrastructure containers. Inclusion is measured against submitted transfers.}
\label{tab:monero-mining-scale}
\end{table}

The Monero experiments run on a server with one AMD EPYC 9654 processor (96 physical cores, 192 logical threads) and approximately 629~GiB of memory. We enable mining on 1, 10, 100, or 200 nodes, with one PoW mining thread on each selected node. Mining nodes are randomly reselected from the full network approximately every 15 minutes, while all nodes continue to maintain the ledger. Each configuration is observed for five hours under a native-transfer workload targeting 0.3~TPS.

We run the 1,200-node network with 200 active single-threaded PoW miners for five hours while rotating miners and processing transfers. The reference chain advances by 324 blocks, and both monitoring nodes ultimately confirm all 5,398 submitted transfers, leaving no pending transactions. All submitted transfers are eventually included across the four configurations. Mean CPU usage rises from 8.858 logical cores with one miner to 186.812 with 200, while mean memory rises from 87.829 to 100.372~GiB (Table~\ref{tab:monero-mining-scale}). CPU consumption grows more sharply than memory use across these deployments.

\paragraph{Solana.}
We also evaluate ChainSEED's Solana service with a 200-node network running Agave, examining whether the deployed clients can sustain consensus progress at this scale. The network uses a low-memory test configuration with approximately five-second slots. During the five-hour observation window, mean container memory consumption is 605.77~GiB and mean CPU usage is 178.08 logical cores, including network infrastructure. Over the same period, the finalized slot number advances by 3,573, showing sustained finalization in this 200-node deployment.

\section{Applications}
\label{sec:applications}
In this section, we demonstrate how ChainSEED supports blockchain research. Researchers may need to modify client behavior, manipulate network conditions, or verify measurements against known connections. Our blockchain services integrate real clients with SEED's configurable network environment, allowing these requirements to be addressed within the same experimental platform. We illustrate these applications through security studies, topology-discovery benchmarking, and network-configuration experiments.

\subsection{Security Studies}
SEED provides topology and routing control, while ChainSEED's services deploy and customize clients for different blockchains. Together, these capabilities let researchers investigate how routing changes affect blockchain operation, how client modifications introduce vulnerabilities, and how attacks behave across blockchain systems.

\paragraph{Routing Attacks: KnockBlock.}
Studying routing attacks against a blockchain requires tracing their effects beyond network reachability to validator duties and chain progress. We illustrate this requirement with KnockBlock, which exploits the interaction between BGP reachability and Ethereum PoS proposer duties~\cite{doumanidis2026routing}. Using our blockchain service, we deployed Geth, Lighthouse, and validator clients in SEED's network to coordinate a routing hijack with a selected validator's scheduled block proposal. From an attacker-controlled network, we announced the network prefix of the beacon node serving this validator, redirecting traffic intended for that node. During the hijack, the validator client could not reach its beacon node, and the assigned proposal was missed. Withdrawing the malicious announcement restored connectivity, demonstrating how ChainSEED supports experiments that link routing manipulation to validator duties.

\paragraph{Blockchain Amplification Attack.}
Security studies may require modifying the blockchain client itself. We illustrate this need with the Blockchain Amplification Attack, which exploits weakened transaction validation to consume client resources without paying for transaction inclusion~\cite{tsuchiya2025amplification}. Our blockchain service automates the deployment of user-provided client binaries. We built Geth from the authors' modified source and integrated it into an emulated Ethereum network through this mechanism. Using the authors' artifact tests, we reproduced the behavior underlying the attack: invalid transactions entered the modified client's transaction pool and displaced honest transactions, yet were excluded from constructed blocks, allowing the attacker to occupy transaction-pool resources without paying for block inclusion. Support for customized clients thus extends the emulator's scope from network-level interventions to security studies of client-internal behavior.

\paragraph{Monero Eclipse Attacks.}
Security studies also extend beyond Ethereum. The experiment reported in Nyx~\cite{shi2026monero} deployed 1,200 Monero nodes across 100 autonomous systems using the Monero service to evaluate an eclipse attack on nodes that do not accept inbound connections. The service configured real clients within SEED's network, enabling controlled connectivity conditions. The experiment showed that these nodes could still be eclipsed, leaving their access to blockchain information under adversarial control.
\subsection{Benchmarking Topology Discovery}
Beyond security studies, our emulator can serve as a benchmark environment for blockchain-network measurement methods. Evaluating topology discovery requires comparing inferred links with a reference topology, yet the actual peer connections are difficult to establish on a public network. Our blockchain service deploys real clients with experimentally controlled peer connections, providing an explicit ground-truth topology against which researchers can evaluate inferred links. We demonstrate this use with TopoShot, which infers Ethereum P2P links through transaction replacement and propagation~\cite{li2021toposhot}.

We extend TopoShot's four-node local validation to different network sizes and connection patterns. Using peer configuration, we construct known topologies on live Geth clients and compare inferred links with reference graphs. Cases cover eight topology families at 4, 6, 8, and 12 nodes, six families at 32 nodes, and a 100-node path. These are measurement-graph sizes. The smaller cases use subgroups of a 12-node deployment with a separate miner.

As shown in Table~\ref{tab:topology-benchmark}, inferred links match the reference graphs in all tested cases with up to 32 nodes. In the 100-node path case, the method identifies 61 of 99 edges, retaining precision of 1.0 but yielding recall of 0.616. With the ground-truth topology available, the benchmark distinguishes missed links from genuinely absent connections, providing a basis for investigating inference errors under different network conditions.
\begin{table}[htbp]
\centering\small
\begin{tabularx}{\linewidth}{@{}Xrrr@{}}
\toprule
Live setting & Cases & Precision & Recall\\
\midrule
4-role local validation & 7 & 1.0 & 1.0\\
4/6/8/12-node subgroups & 32 & 1.0 & 1.0\\
32-node topologies & 6 & 1.0 & 1.0\\
100-node path & 1 & 1.0 & 0.616\\
\bottomrule
\end{tabularx}
\caption{Topology-discovery precision and recall on known peer graphs.}
\label{tab:topology-benchmark}
\end{table}

The same benchmark environment could support related topology-discovery methods such as DEthna~\cite{zhao2024dethna}. Researchers can configure peer topologies and deploy their own probing algorithms to examine how different connection patterns affect their methods, using the resulting feedback to refine algorithm design before applying them to topology discovery on public networks.

\subsection{Network Protocol Configuration Studies}
Beyond discovering how blockchain nodes are connected, researchers may ask how those connections should be configured. Maintaining more peers provides additional paths for consensus communication, but also increases the traffic each node handles. Choosing a configuration therefore requires understanding whether the extra communication brings a corresponding benefit to information dissemination. Our emulator allows researchers to examine this trade-off by varying client settings within the same underlying network and observing both traffic and chain progress. Motivated by Eth2.0-NA's study of forwarding redundancy~\cite{zhao2026eth20na}, we investigate this question through Lighthouse's target peer count~\cite{lighthouse-networking}.

To examine whether fewer connections can reduce traffic while preserving timely chain updates, we deploy ten Lighthouse beacon nodes and their validator clients alongside one Geth execution node in a multi-AS network. With the underlying topology and other settings unchanged, we compare target peer counts of eight and seven, then reduce the target to five to explore a further reduction. For each configuration, we observe network traffic and the spread of beacon-head observations over a 180-second window.

As shown in Table~\ref{tab:network-parameter-study}, reducing the target peer count from eight to seven lowers measured traffic by 3.75\%. Reducing it to five yields a 19.04\% reduction relative to eight. Meanwhile, the mean beacon-head observation spread remains close to 0.94 seconds across all three configurations. In this deployment, fewer peer connections therefore reduce communication cost without a discernible increase in the sampled observation spread. Comparing these outcomes allows researchers to assess the communication benefit of additional connections against their traffic cost.

\begin{table}[htbp]
\centering\small
\begin{tabularx}{\linewidth}{@{}r*{2}{>{\centering\arraybackslash}X}@{}}
\toprule
Target peers & Aggregate traffic (bytes) & Mean observation spread (s)\\
\midrule
8 & 23,637,589 & 0.9378\\
7 & 22,750,408 & 0.9379\\
5 & 19,137,784 & 0.9336\\
\bottomrule
\end{tabularx}
\caption{Beacon network traffic and chain-head observation spread.}
\label{tab:network-parameter-study}
\end{table}

\section{Discussion and Future Work}
\label{sec:discussion-future-work}

\subsection{Discussion}
Scaling a blockchain emulator requires choosing which behaviors an experiment must preserve within a hardware budget. Our Ethereum results show that the appropriate resource configuration differs across client roles. Lower beacon concurrency saves memory but can increase queueing, while a larger execution-client memory budget can improve transaction processing. Validator multiplexing reduces repeated process overhead while retaining individual identities and consensus duties. Researchers can use these choices to allocate resources toward the node network, transaction workload, or validator population required by their study. The resulting scale should be assessed together with chain synchronization, transaction completion, and consensus progress.

The grouping of clients also shapes the conditions an experiment can represent. A study of communication among consensus nodes needs independently deployed participants and their peer connections. A study of validator duties can retain a fixed identity population while varying the clients that host it. If the question concerns client failures, that grouping becomes part of the experimental design: a failure in a shared validator client affects every identity it manages. Researchers can choose the multiplexing degree according to both available resources and the intended scope of a failure. Comparisons can then distinguish the failure of one client from the loss of one validator identity.

Client and network controls let researchers change conditions relevant to a hypothesis and observe the response of the running chain. In a transaction-load experiment, execution-client transaction pools and beacon processing queues reveal different kinds of waiting, while block inclusion and finalization indicate how the chain progresses. In the routing-attack study, the missed proposal is examined together with the validator's access to its beacon node and changes in network reachability. Known peer connections provide a reference for evaluating topology-discovery results and investigating missing links. These observations help researchers decide whether to change a client setting, a workload, or the network environment in a follow-up experiment. The same chain definition can be reused when comparing client placements or peer configurations.

\subsection{Future Work}

\paragraph{Ethereum Layer 2.}
Ethereum's rollup-based scaling strategy motivates experiments that connect Layer-2 transaction processing with Layer-1 data publication and settlement~\cite{ethereum-scaling}. We plan to deploy rollup clients and their Layer-1 contracts within the same emulated network. Researchers could vary workloads and network conditions across both layers to examine how sequencer availability and Layer-1 congestion affect throughput and settlement latency. For example, a fixed Layer-2 workload could be run under different Layer-1 loads to examine where transactions wait before settlement. Separately changing sequencer reachability would allow network effects to be compared with the effects of transaction demand. These experiments would also measure the resources required to host both layers and assess the deployment scales that fit a shared hardware budget.

\paragraph{Monero and Solana Optimization.}
Monero and Solana place different demands on the resources available to an experiment. We plan to examine how their client configurations affect memory and CPU consumption when scaling node populations or transaction loads. For Monero, this includes the cost of maintaining a ledger alongside concurrent PoW mining, varying active miners separately from the node population. Separately tracking active miners and non-mining nodes would help identify which part of the deployment benefits from a configuration change. For Solana, we will study transaction processing and finalization costs under different workloads. Each configuration will be evaluated together with transaction completion and chain progress to identify useful resource savings for larger deployments.

\paragraph{Validator Consolidation.}
EIP-7251, activated with Ethereum's Pectra upgrade on May 7, 2025, raises the maximum effective validator balance from 32 to 2,048~ETH while retaining the 32~ETH activation minimum, enabling opt-in consolidation~\cite{ethereum-pectra,eip-7251,ethereum-maxeb}. Client multiplexing preserves validator identities, while consolidation combines stake into fewer identities. We plan to support variable balances and consolidation in ChainSEED and study their interaction with client multiplexing. Holding total effective stake fixed, researchers could vary both the number of identities and the number of clients hosting them. This would allow resource use, duty execution, and consensus progress to be examined as consequences of identity consolidation and client grouping. The comparison could also track the effective stake affected by a client failure as balances and client groupings change.

\paragraph{Client Diversity.}
We plan to integrate execution and consensus clients beyond Geth and Lighthouse and compare them under matched network conditions, protocol settings, and transaction workloads. Using each client's configuration options, we will examine resource consumption, transaction processing, and consensus progress to assess whether optimization findings transfer across implementations. Comparing resource costs at similar levels of throughput and consensus performance would help researchers choose client implementations and configurations for their available hardware budget.

\section{Conclusion}
\label{sec:conclusion}
ChainSEED provides researchers with a configurable experimental platform built from real blockchain clients. Its client and network controls make it possible to connect changes in software, peer relationships, and workloads to observed transaction processing and consensus behavior. Our studies show that useful experimental scale depends on allocating resources to the behaviors under investigation, and demonstrate how the resulting environment supports security studies and network measurement. Researchers can use ChainSEED to reproduce phenomena, test mechanisms and applications, and examine why observed behavior differs from their expectations. This gives blockchain research a setting where hypotheses can be investigated through controlled interventions in a running system.

\FloatBarrier
\bibliography{design-references}

\FloatBarrier
\appendix
\section{Existing Software Workflows}
\label{sec:software-fidelity}
Deploying a large blockchain network provides the setting for an experiment, but researchers also need to interact with it through the tools they use to develop and test blockchain applications. These activities range from wallet transfers and interactive contract calls to automated deployment and testing. Building on the interfaces described in Section~\ref{sec:software-integration}, we evaluated these workflows using MetaMask, Remix, and Hardhat connected to the clients running in ChainSEED.

\begin{figure}[!htb]
\centering
\begin{minipage}[t]{\linewidth}\centering
\includegraphics[width=\linewidth]{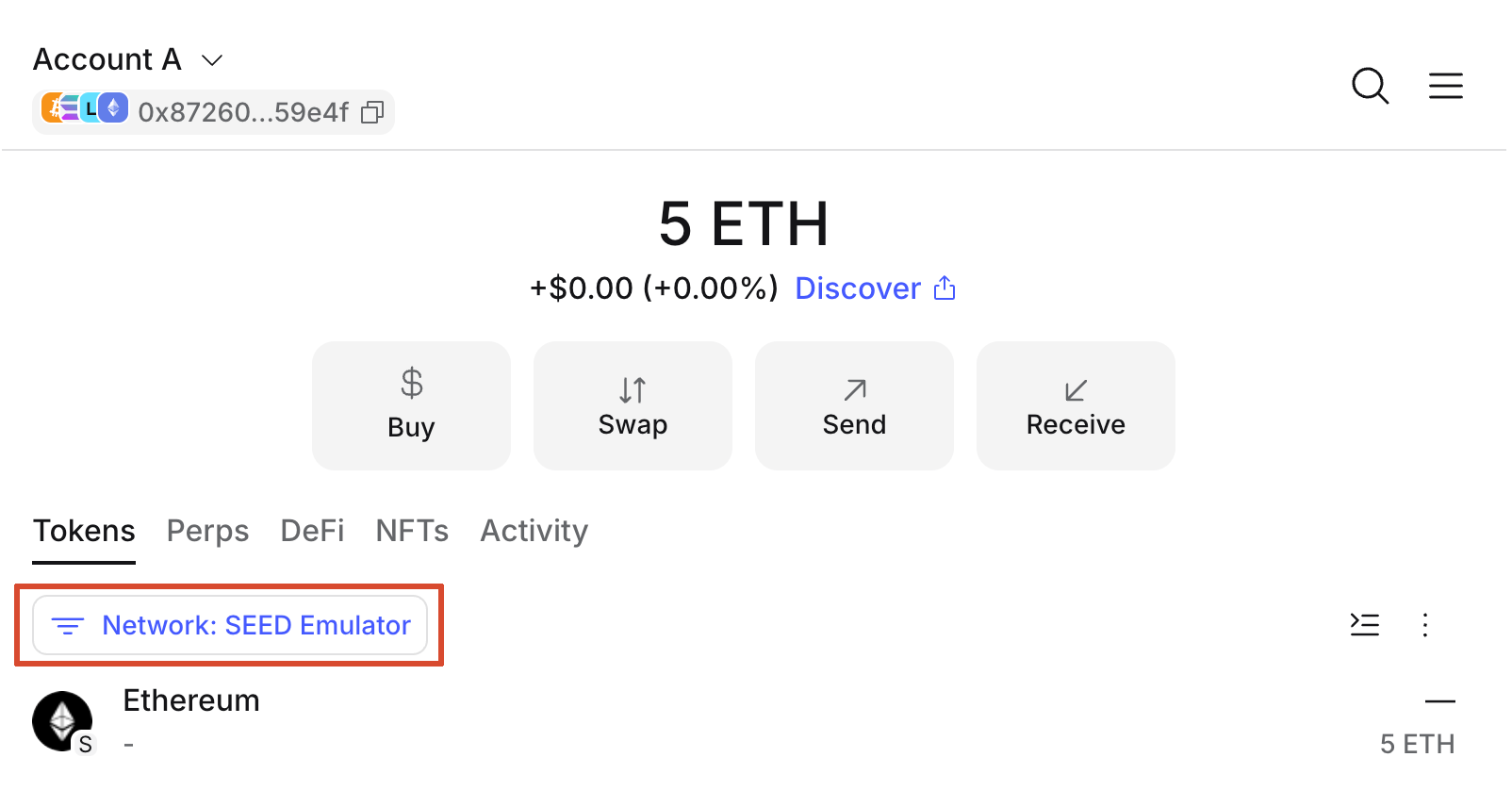}
\end{minipage}\par\smallskip
\begin{minipage}[t]{\linewidth}\centering
\includegraphics[width=\linewidth]{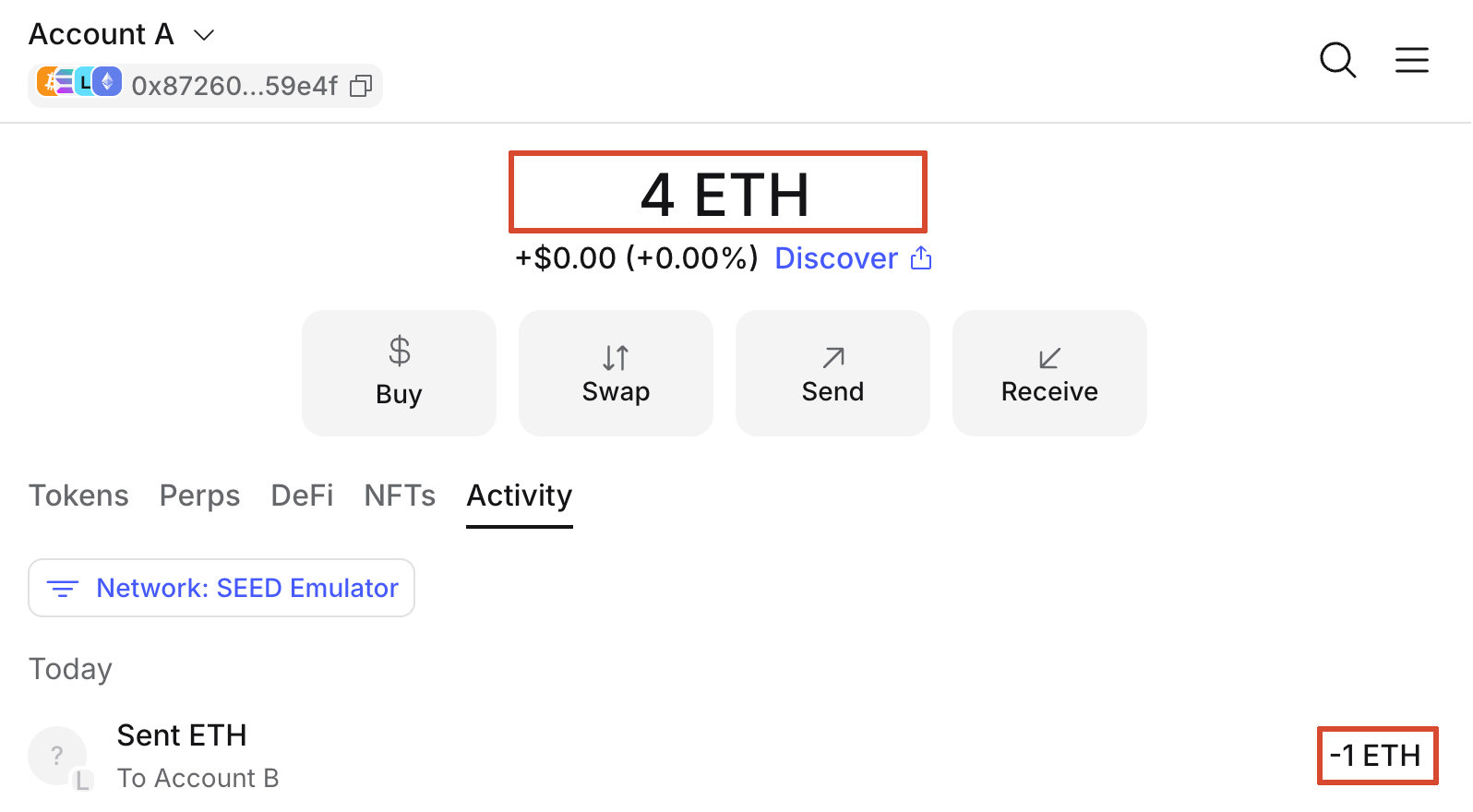}
\end{minipage}
\caption{Wallet interaction through MetaMask.}
\Description{Two MetaMask screenshots show Account A on the SEED Emulator network, first with a displayed balance of 5 ETH and then 4 ETH with an activity entry for sending 1 ETH to Account B.}
\label{fig:software-demonstration}
\end{figure}

\begin{figure*}[!t]
\centering
\includegraphics[width=0.75\linewidth]{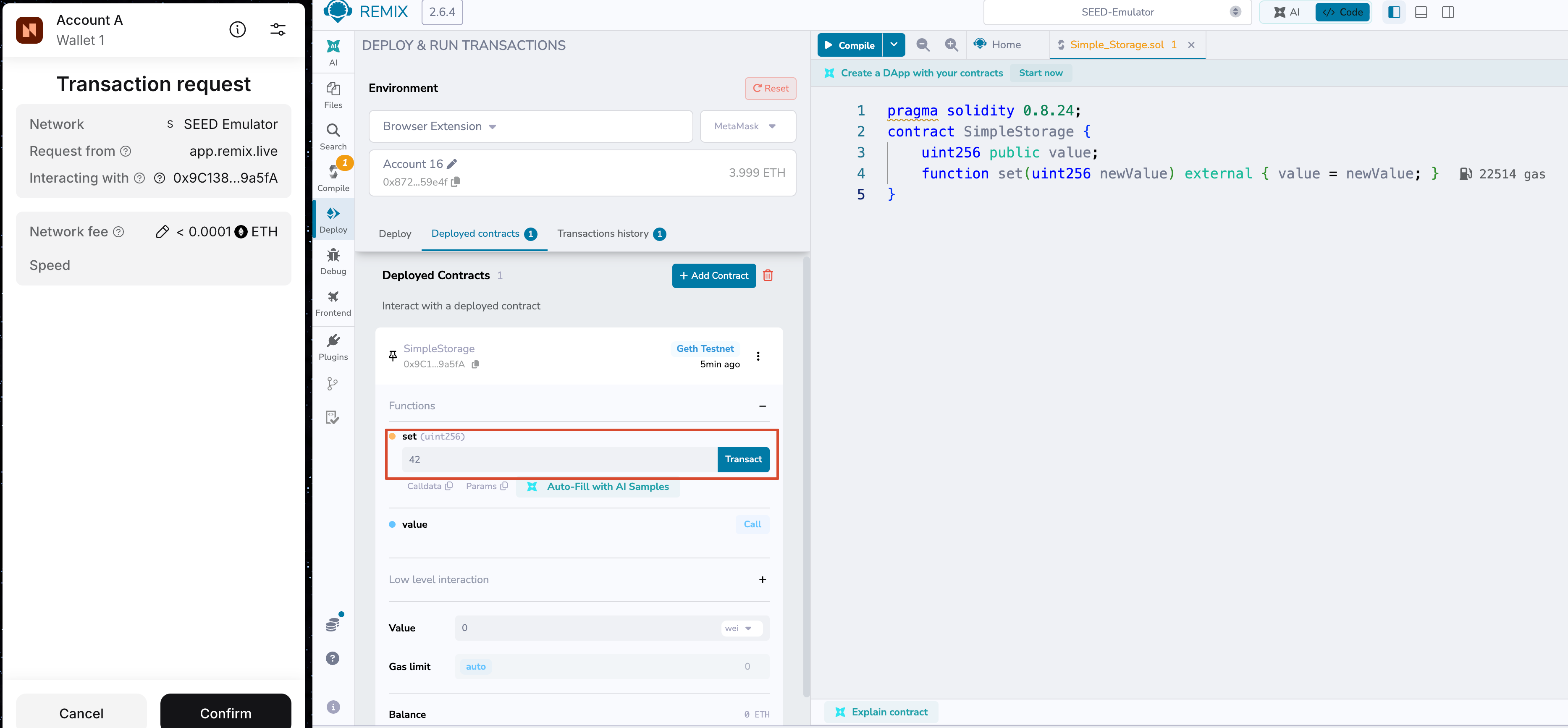}
\caption{Contract interaction through Remix and MetaMask.}
\Description{Remix displays the SimpleStorage contract, the browser-extension environment, and the set function with input 42. The adjacent MetaMask window shows a transaction request from Remix on the SEED Emulator network.}
\label{fig:remix-interaction}
\end{figure*}

We began with a basic wallet interaction: transferring 1~ETH from Account A to Account B through MetaMask. The transaction completed successfully on the emulated chain. We used Remix to deploy a simple contract that stores an integer and allows users to update and query its value. We then submitted a contract call through MetaMask to change the stored value from 0 to 42 and verified the updated value in Remix. Figures~\ref{fig:software-demonstration} and~\ref{fig:remix-interaction} illustrate the wallet and contract interfaces used in these experiments.

\begin{figure*}[!t]
\centering
\includegraphics[width=0.75\linewidth]{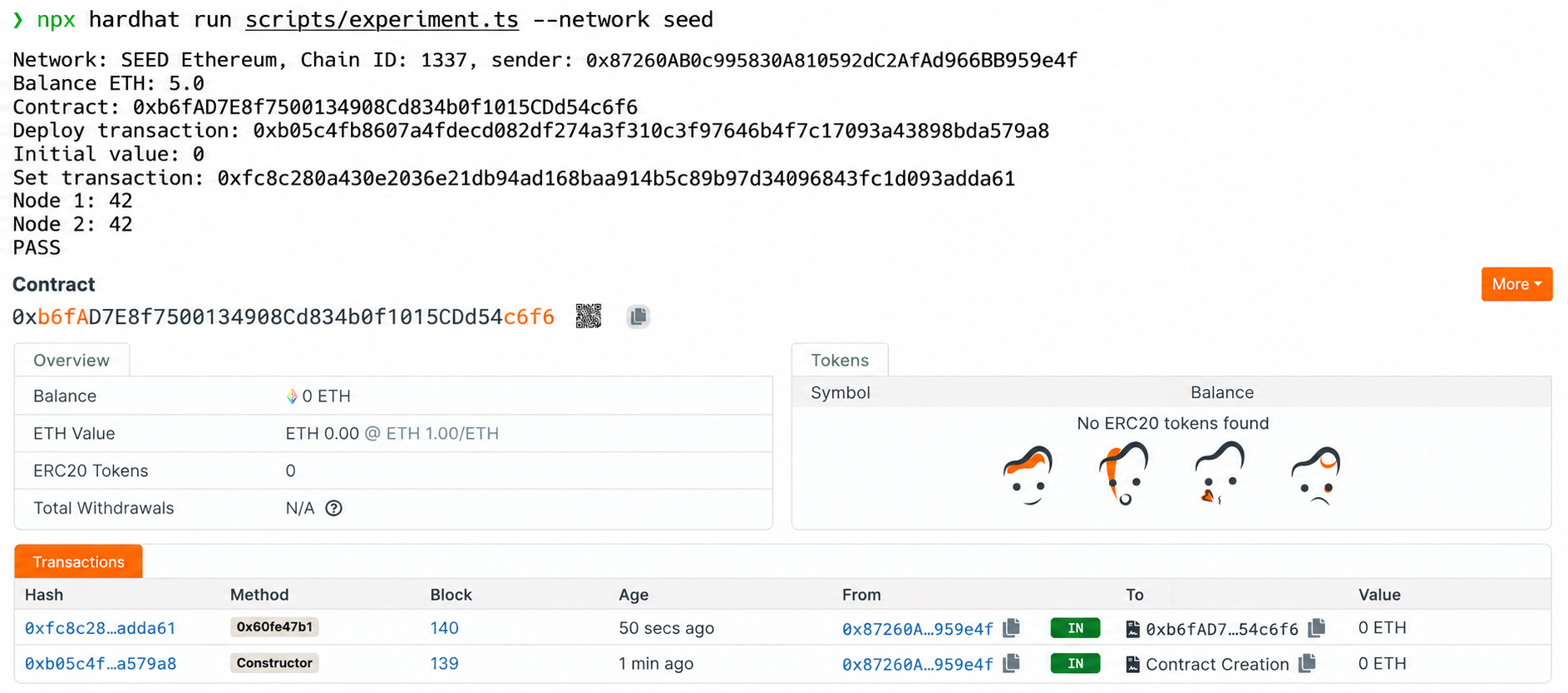}
\caption{Hardhat contract interaction and its on-chain records.}
\Description{The Hardhat terminal output reports chain ID 1337, initial contract value zero, updated values of 42 from two nodes, and PASS. The explorer lists the matching contract creation and state-changing transactions in blocks 139 and 140.}
\label{fig:hardhat-interaction}
\end{figure*}

We used Hardhat to deploy the same integer-storage contract and update its value from 0 to 42 through the emulated chain's RPC interface. The script then queried two execution nodes, both of which returned the updated value. Figure~\ref{fig:hardhat-interaction} shows the script output alongside EthExplorer's records of the deployment and contract call.

Together, these experiments show that users can carry familiar wallet, contract-development, and automated testing workflows into ChainSEED. Transactions and contract calls are processed by the deployed clients, while the existing tools retain their usual interfaces. Researchers can therefore develop and test applications within a configurable blockchain environment using the software they already use.

\section{EthExplorer}
\label{sec:ethexplorer-views}
\begin{figure*}[!t]
\centering
\includegraphics[width=0.75\linewidth]{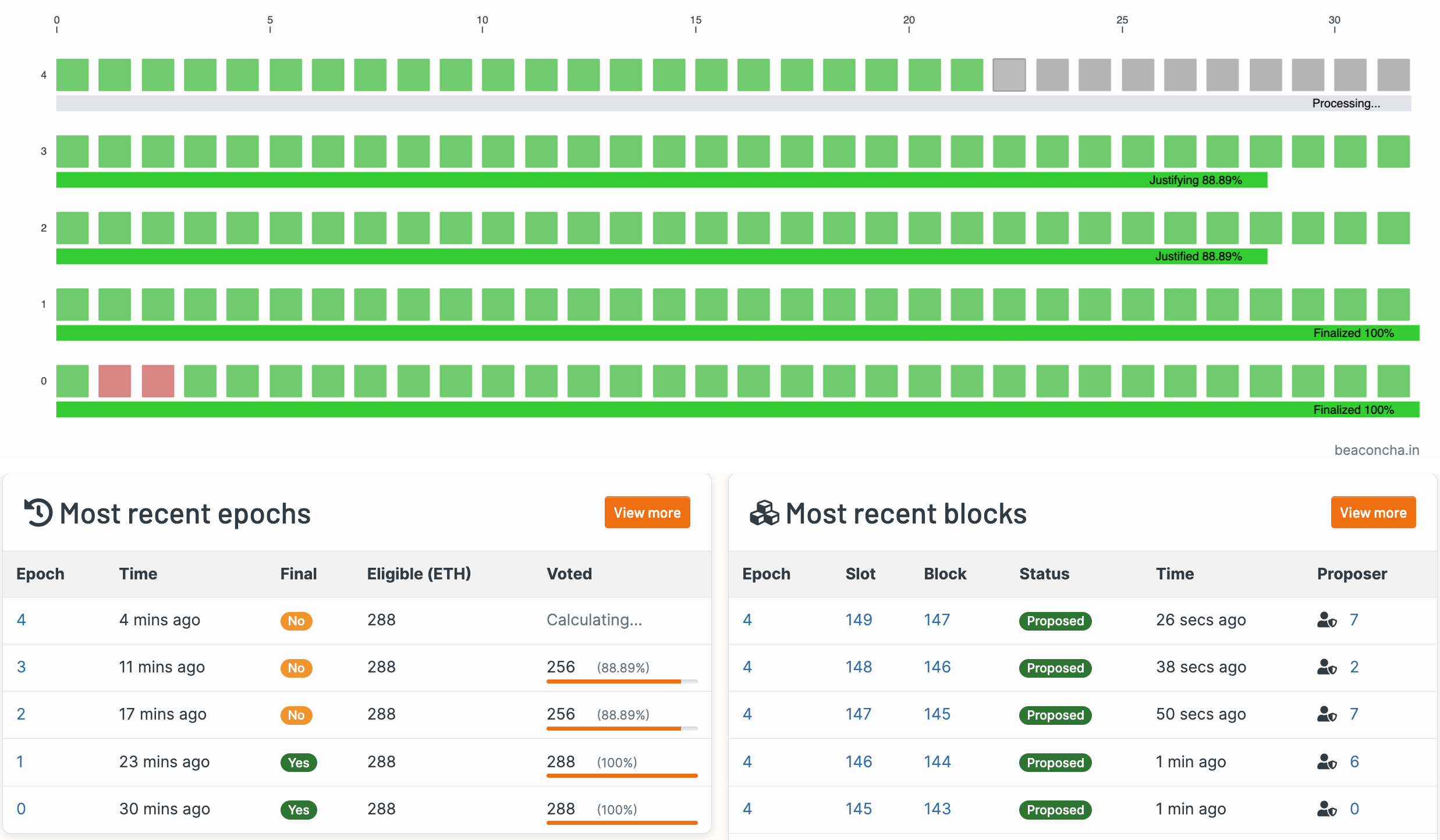}
\caption{Chain activity viewed through EthExplorer.}
\Description{Two stacked EthExplorer screenshots show slot status across epochs zero through four, justification and finalization, and tables of recent epochs and proposed blocks.}
\label{fig:ethexplorer-view}
\end{figure*}

Figure~\ref{fig:ethexplorer-view} illustrates the chain-level view provided by EthExplorer, described in Section~\ref{sec:monitoring-system}. The view presents an epoch and slot overview alongside recent epochs and blocks.

\section{Ethereum Lookingglass}
\label{sec:lookingglass-views}
\begin{figure*}[!t]
\centering
\includegraphics[width=0.75\linewidth]{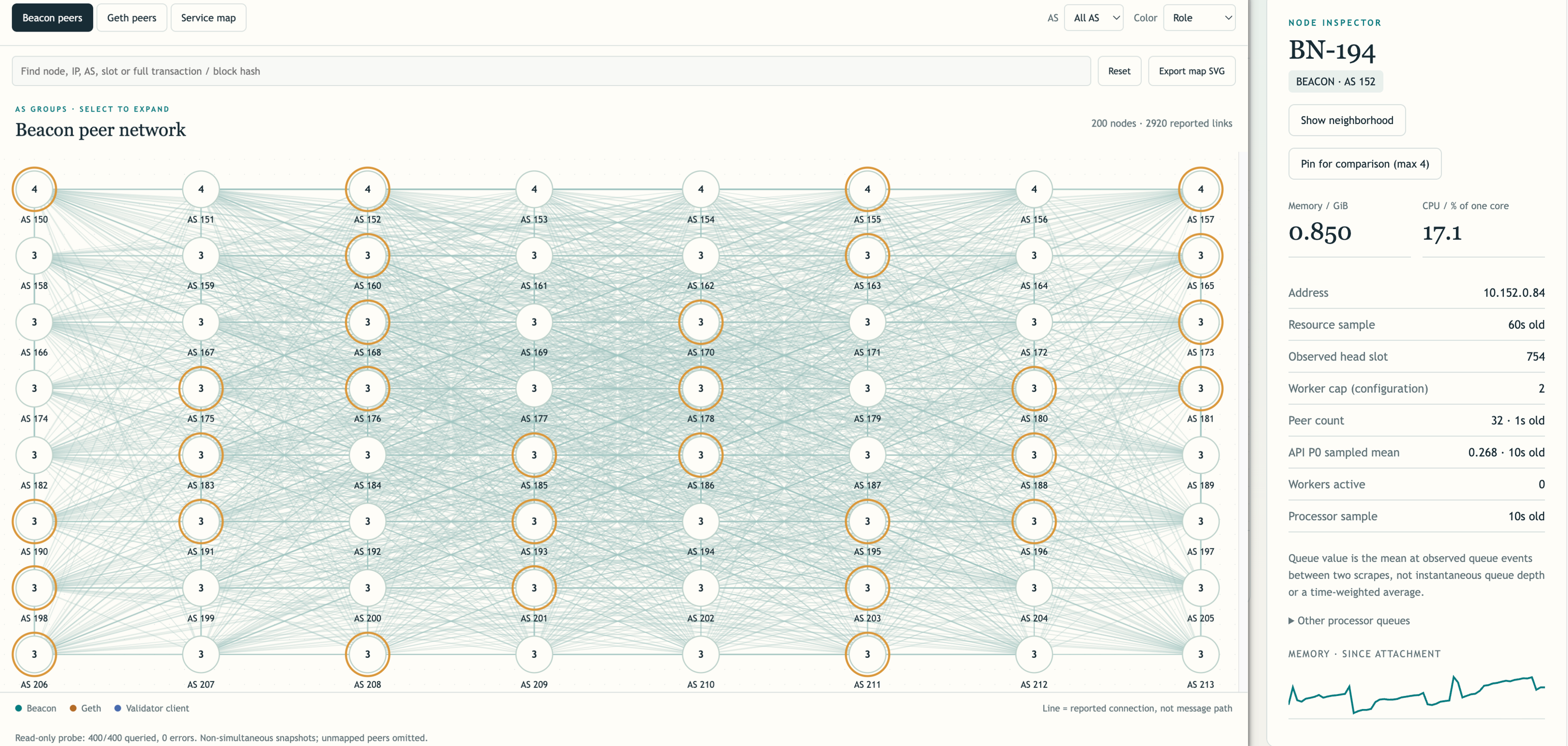}
\caption{Peer connections and node behavior in Ethereum Lookingglass.}
\Description{Screenshot of Ethereum Lookingglass with the All AS view selected. AS groups summarize 200 Beacon nodes and 2920 reported links. The inspector for BN-194 shows memory and CPU usage, observed head slot, configured worker cap, peer count, sampled API P0 queue mean, active workers, and a memory trend.}
\label{fig:lookingglass-view}
\end{figure*}

Figure~\ref{fig:lookingglass-view} illustrates the node-level view provided by Ethereum Lookingglass, described in Section~\ref{sec:monitoring-system}. The view shows beacon peer connections grouped by AS alongside the selected node's resource usage and processor metrics.

\section{Transaction Workload Composition}
\label{sec:transaction-mix-appendix}
Table~\ref{tab:transaction-mix} lists the default category sampling weights used by the transaction generator described in Section~\ref{sec:transaction-generation}. These weights specify sampling probabilities rather than exact proportions enforced in each interval.

\begin{table}[htbp]
\centering\small
\begin{tabular}{lr}
\toprule
Transaction category & Sampling weight (\%)\\
\midrule
Native ETH transfer & 31.6839\\
ERC-20 transfer & 30.4803\\
Generic medium-cost call & 14.0826\\
Generic heavy call & 9.5467\\
Generic light call & 6.2499\\
ERC-20 approve & 3.5743\\
DEX swap & 2.2243\\
transferFrom & 1.4688\\
NFT operation & 0.4113\\
Account abstraction & 0.1969\\
Contract deployment & 0.0811\\
\bottomrule
\end{tabular}
\caption{Sampling weights for transaction categories.}
\label{tab:transaction-mix}
\end{table}

\end{document}